\documentclass[journal]{IEEEtran}
\usepackage{tikz}
\usepackage{amssymb}  
\usepackage{amsmath}
\usepackage{booktabs}
\usepackage{subcaption}

\newtheorem{theorem}{Theorem}{\bf}{}
\newtheorem{problem}{Problem}{\bf}{}
{\bf}{}
{\bf}{}

\ifCLASSINFOpdf
\else
\fi
\begin{document}
%

\title{Mean-field Games for Bio-inspired Collective Decision-making in Dynamical Networks}

%
%
%


\author{Leonardo~Stella and Dario~Bauso
\thanks{Additional contributions with respect to the conference version \cite{SB17M} include: i) the study of the microscopic dynamics of the system for a finite number of players interacting according to a specific topology, ii) the specialization of the model to describe honeybee swarms,  virus propagation, and cascading failures in smart-grids, and iii) the numerical analysis of a set of cyber-attacks in the context of smart grids.}
\thanks{L. Stella is with Department of Automatic Control and Systems Engineering, The University of Sheffield, Mappin Street Sheffield, S1 3JD, United Kingdom
       e-mail: (lstella1@sheffield.ac.uk)}%
\thanks{D. Bauso is with Department of Automatic Control and Systems Engineering, The University of Sheffield, Mappin Street Sheffield, S1 3JD, United Kingdom, and Dipartimento di Ingegneria Chimica, Gestionale, Informatica, Meccanica,  Universit\`a di Palermo, V.le delle Scienze, 90128 Palermo, Italy
        e-mail: (d.bauso@sheffield.ac.uk)}%
}

\maketitle

\begin{abstract}                
Given a large number of homogeneous players that are distributed across three possible states, we consider the problem in which these players have to control their transition rates, while minimizing a cost. The optimal transition rates are based on the players' knowledge of their current state and of the distribution of all the other players, and this introduces mean-field terms in the running and the terminal cost. The first contribution involves a mean-field game model that brings together macroscopic and microscopic dynamics. We obtain the mean-field equilibrium associated with this model, by solving the corresponding initial-terminal value problem. We perform an asymptotic analysis to obtain a stationary equilibrium for the system. The second contribution involves the study of the microscopic dynamics of the system for a finite number of players that interact in a structured environment modeled by an interaction topology. The third contribution is the specialization of the model to describe honeybee swarms, virus propagation, and cascading failures in interconnected smart-grids. A numerical analysis is conducted which involves two types of cyber-attacks. We simulate in which ways failures propagate across the interconnected smart grids and the impact on the grids frequencies. We reframe our analysis within the context of Lyapunov's linearisation method and stability theory of nonlinear systems and Kuramoto coupled oscillators model.

\end{abstract}


\IEEEpeerreviewmaketitle

\section{Introduction}
We consider a large population of players that choose any of three possible states, state \emph{1}, state \emph{2} and state \emph{3}. These states can be also seen as opinions. We model the problem as a continuous-time discrete-state mean-field game. Players transition from one state to another according to  transition rates that are influenced by adversarial disturbances. Thus, the evolution of the state of a player follows a traditional Markov chain dynamics with controlled and uncontrolled transition rates. Players control their transition rates to minimize a cost functional defined over a finite horizon. The cost functional is composed by a running cost and a terminal penalty. The running cost accounts for the actual transition rate from a state to another and an additional term which is a function of the state. Depending on the function, our model includes both crowd-seeking and crowd-averse behaviours. The terminal penalty has the same structure of the running cost in that it weighs the final state of the player at the end of the horizon. We further assume that the cost of transitioning from state \emph{1} to state \emph{2} is sufficiently large to discourage direct transitions from \emph{1} to \emph{2} and vice versa. We assume that players are homogeneous.\\

\noindent \textbf{Highlights of contributions.} 
The main contribution of this paper is threefold. First, we build a mean-field model where the macroscopic and microscopic dynamics are brought together in a unified framework. 
Second, we provide mean-field response strategies for the players and the disturbances. Third, we study stationary mean-field equilibrium points and prove that such stationary solutions can be obtained in the asymptotic limit of nonstationary mean-field equilibrium points, for which we provide an explicit expression. 
The following is a list of additional results with respect to the conference paper, see \cite{SB17M}: i) we extend our study to the case of honeybee swarm and virus propagation in the framework of cyber-security; ii) we model the microscopic dynamics with a finite number of players, which interact according to a specific topology. The state of the model is the probability that any given player can be in each of the three possible states; iii) we provide a case study for a UK power grid, where we study two types of cyber-attacks; the underlying Markov model simulates the propagation of the failure due to the cyber-attack and the impact on the grid frequency; this simulation is based on the Kuramoto coupled oscillators model.\\

\noindent \textbf{Related literature}. The origin of the theory of mean-field games is in \cite{HCM03}, \cite{HCM06}, \cite{HCM07} by M. Y. Huang, P. E. Caines and R. Malham\'e and independently in \cite{LL06a}, \cite{LL06b}, \cite{LL07} by J. M. Lasry and P. L. Lions. In the context of Markov decision processes, the authors in \cite{WBV05} introduce a concept similar to the \emph{mean-field equilibrium}, called \emph{oblivious equilibrium}. For a survey on mean-field games we refer the reader to \cite{GSa}. Robust mean-field games are introduced in \cite{BTB} and studied further in \cite{DTA}. Robustness is also discussed in \cite{TZBa}.\\
Mean-field games apply to a variety of domains, including economics, engineering, physics and biology; for details we refer the reader to \cite{HCM07}, \cite{ACC12}, \cite{BagBau13}, \cite{GLL10}, \cite{PB11}, \cite{TBB13}. Mean-field games have predecessors in \emph{anonymous games} and \emph{aggregative games} where the notion of mass interaction was already a main feature. Examples of finite state mean-field games can be found in \cite{Gomes}. For a linear quadratic structure, the author in \cite{B12} provides explicit solutions in terms of mean-field equilibria. Otherwise, numerical approximations and discretization methods can be used as one of several solution schemes proposed in recent times, see \cite{ACC12}. \\
Lately, cyber-security and prevention/protection mechanisms against virus propagation and cyber-attacks have sparked increasing interest. In the context of smart grids, cyber-attacks in the form of of data attacks are studied in \cite{KJTT11}. Epidemic models to study failure propagation is proposed in \cite{CB16}. One key mechanism to contrast these attacks is to provide interdependency between critical infrastructures. Such infrastructure can be physical, logical or cyber, see \cite{RP17}.
The novelty of this work with respect to the discussed literature is the development of a mean-field game to accomodate existing models of bio-inspired collective decision-making and provide for them an evolutionary game interpretation. We also extend it to cyber-security in the scenarios where a failure induced by a cyber-attack propagates in the same ways in which viruses propagate in a network. Two different virus propagation models are considered: the Susceptible-Infected-Recovered (SIR) and the Susceptible-Infected-Susceptible (SIS) models. The first model describes the human immunity system in relation to the diffusion of specific viruses such as measles, mumps and rubella, while the second model extends to viruses that can be contracted again after healing such as in patients affected by x-linked agammaglobulinemia, who are unable to develop immunities to common diseases and infections. The impact of cyber-attacks on the grid is measured in terms of frequency oscillations; the simulation model builds on the analogy with Kuramoto coupled oscillators, see~\cite{FBLN}. \\

This paper is organised as follows. In Section~\ref{sec:mfm} we define the mean-field model. In Section~\ref{sec:mfg} we study  equilibrium points and stability. In Section~\ref{sec:hs} we introduce the honeybee swarm model. In Section~\ref{sec:vp} we study the virus propagation network scenario.
In Section~\ref{NA} we provide numerical analysis and a case study of a real UK power grid. In Section~\ref{sec:conc} we provide conclusions and future works.

\section{Mean-field Game Model}\label{sec:mfm}
In this section, a mean-field model for a decision-making problem with three possible choices is presented. First, a general formulation of the problem is given for the macroscopic model. Then, the perspective of a reference player is tackled: the optimal control problem is analysed and finally the mean-field response for the reference player is presented.

\subsection{Macroscopic model: Kolmogorov equations}
\begin{figure}[h]
\begin{centering}
      \begin{tikzpicture}
      \path 	(0,0) node[circle,draw](x) {$3$}
      		(-3,1) node[circle,draw] (z) {$1$}
		(3,1) node[circle,draw](y) {$2$};
	\draw[->,black] (x) .. controls +(up:1cm) and +(right:1cm) .. node[above] {$\beta_{31}$} (z);
	\draw[->,black] (z) .. controls +(down:1cm) and +(left:1cm) .. node[below] {$\beta_{13}$} (x);
	\draw[->,black] (x) .. controls +(up:1cm) and +(left:1cm) .. node[above] {$\beta_{32}$} (y);
	\draw[->,black] (y) .. controls +(down:1cm) and +(right:1cm) .. node[below] {$\beta_{23}$} (x);
      \end{tikzpicture}
\caption{Markov chain of the system.}
\label{fig:markov}
\end{centering}
\end{figure}
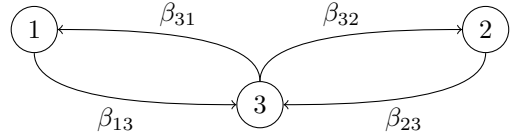

We consider a large population of players being in any of three possible states in a continuous-time dynamic game framework. These players control their states, according to some optimality criteria. We further assume that all players behave in the same way, i.e. they are homogeneous. The game is symmetric with respect to any permutation of players, i.e. the decision of each player does not take into account the individual players but rather the distribution of players in each state. The controls of players depend on the knowledge of their own position and of the distribution of the population among the possible states. We assume that the state of a player evolves following the Markov chain depicted in Fig. \ref{fig:markov}, where $\beta_{ij} \in \mathbb{R}^{3\times 3}$ represents the transition rate from node $i$ to node $j$, for any generic pair of nodes $i,j \in \mathbb{I}^3$, where $\mathbb{I}^3$ is the set of three possible states, state $1$, state $2$ and state $3$. Nodes $1$ and $2$ are the are two alternative choices a player can commit to, while node $3$ is the uncommitted state.

We tackle the problem from the perspective of a single player, hereafter referred to as the \emph{reference player}.. We assume that the only information available to the reference player is the distribution of the whole population modeled by the probability vector $\mathbf{x}(t) = [x_1, x_2, x_3]^T \in \mathcal{S}^3$, where $\mathcal{S}^3$ is the probability simplex in $\mathbb{R}^3$ and $t>0$ is the time index. Players change state according to a continuous time Markov process with transition rate matrix $\beta(t) \in \mathbb{R}^{3\times 3}$. Each column of the matrix $\beta$ has two components, i.e. $\beta_i = \rho_i + w_i$, where the first component $\rho_i \in (\mathbb{R}^+_0)^3$ is controlled by the player and the second $w_i \in (\mathbb{R}^+_0)^3$ is controlled by an adversarial disturbance. Each component of the matrix $\beta$, i.e. $\beta_{ij}$, represents the transition rate from node $i$ to node $j$, as previously mentioned. The three possible states are: 
\begin{itemize}
\item state $1$: committed to opinion $A$;
\item state $2$: committed to opinion $B$;
\item state $3$: uncommitted.
\end{itemize}
In the mean-field limit when the number of players tends to infinity, the model is described by the following Kolmogorov equations:
\begin{equation} \label{eq:um}
\begin{array}{ll}
\dot{x}_1 = x_3\beta_{31} - x_1\beta_{13}, \\
\dot{x}_2 = x_3\beta_{32} - x_2\beta_{23}, \\
\dot{x}_3 = x_1\beta_{13} + x_2\beta_{23} - x_3\beta_{31} - x_3\beta_{32}.
\end{array}
\end{equation}
This model has an initial condition for the distribution $\mathbf{x}(0) = \mathbf{x}_0$. From the conservation of mass, we know that $\dot{x}_3 = 1 - x_1 - x_2$, thus equation (\ref{eq:um}) can be rewritten as:
\begin{equation} \label{eq:um2}
\begin{array}{ll}
\dot{x}_1 = (1 - x_1 - x_2)\beta_{31} - x_1\beta_{13}, \\
\dot{x}_2 = (1 - x_1 - x_2)\beta_{32} - x_2\beta_{23}. \\
\end{array}
\end{equation}
Then the macroscopic model is represented by the above set of nonlinear ODEs in the bi-dimensional state space $x_1$-$x_2$. 

\subsection{Microscopic model: mean-field response}
Let us consider the reference player and model their microscopic dynamics under the assumption that the distribution over the time horizon is given. The state of the player indicates which opinion it is committed to, or if it is uncommitted, and thus the state takes value in a finite discrete set with cardinality 3. Let the state of the player be denoted by $i \in \mathbb{I}^3$. The evolution of the state is described using a continuous-time Markov chain. The transition rates are chosen to minimise a total cost consisting of a running cost and a terminal cost.
The running cost depends on the state of the player, the distribution of the population and the transition rate, i.e. $g(i, x_i, \rho_i, t): \mathbb{I}^3 \times \mathcal{S}^3 \times (\mathbb{R}^+_0)^3 \times \mathbb{R}^+_0 \rightarrow \mathbb{R}$ is defined as:
\begin{equation} \label{eq:g}
g(i, x_i, \rho_i, t) =  \frac{1}{2}\|\rho_i\|^2_{R_i} + f_i(x_i(t)),
\end{equation}
where $\rho_i \in (\mathbb{R}^+_0)^3$ is the transition rate for the reference player, $\|\rho_i\|^2_{R_i}$ is the second norm weighted on matrix $R_i$ and $R_i \in \mathbb{R}^{3\times3}$ is a $3\times3$ diagonal matrix, defined as $R_i=diag([R_{i1},R_{i2}, R_{i3}])$, where we denote by $diag(a)$ a diagonal matrix with diagonal $a$ for any generic vector $a$. 
We have chosen $g$ such that:
\begin{itemize}
\item it is Lipschitz continuous in $x_i$, with the Lipschitz constant with respect to $x_i$ bounded independently of $\rho_i$,
\item and it is differentiable with respect to $\rho_i$, and the derivative with respect to $\rho_i$ is Lipschitz continuous with respect to $x_i$, $\forall i \in \mathbb{I}^3$.
\end{itemize}
It can be verified that $g(i, x_i, \rho_i, t)$ does not depend on $\rho_i$, is uniformly convex and is superlinear on $\rho_j$, $j\neq i$.\\
Let us consider a finite horizon formulation of the game and use $[0,T]$ to indicate the horizon window. Thus, the reference player also incurs in a terminal cost depending on the objective they seek to minimize. Let the terminal cost be $\psi(i, x_i): \mathbb{I}^3 \times \mathcal{S}^3 \rightarrow \mathbb{R}$, defined as:
\begin{equation} \label{eq:psi}
\psi(i, x_i) = f_i(x_i(T)).
\end{equation}
Then, each player minimises the cost functional 
\begin{equation} \label{eq:J}
\begin{array}{lll}
J_{x_i}^i(\rho_i, w_i, t) \\ 
\quad = \mathbb{E}_{{i_t} = i}^{\rho_i, w_i}\Bigg[\int_t^T \Big[g(i_{\tau}, x_{i_{\tau}}(\tau), \rho_{i_{\tau}}(\tau)) - \frac{1}{2}\|w_{i_{\tau}}\|^2_{\Gamma_{i_{\tau}}}\Big] d\tau \\
\quad + \psi(i_T, x_{i_T})\Bigg],
\end{array}
\end{equation}
where $i_{\tau}$ is the continuous-time Markov chain and describes the state of the reference player at time $\tau$, the term $\|w_i\|^2_{\Gamma_i}$ indicates the second norm weighted by matrix $\Gamma_i$ of vector $w_i$, $\Gamma_i \in \mathbb{R}^{3\times3}$ is a $3\times3$ diagonal matrix defined as $\Gamma_i=diag([\Gamma_{i1},\Gamma_{i2}, \Gamma_{i3}])$,
and $\mathbb{E}_{i_t = i}^{\rho_i, w_i}$ is the expectation for the event $i_t = i$.
The term in $w_i \in (\mathbb{R}_0^+)^3$ represents a penalty on the energy of the disturbance signal. The functional has the structure of a robust mean-field game in spirit with $H_\infty$-optimal control, see \cite{BTB}. Roughly speaking, increasing entries for $\Gamma_i$ leads to smaller disturbances in the second norm.\\

\begin{problem}\label{prob1}
Let $\mathbf{x}(t): [0, T] \rightarrow \mathcal{S}^3$ be given. Find the optimal control of the reference player which minimises the cost functional:
\begin{equation} \label{eq:v}
v_i(x_i,t) = \inf_{\rho_i(\cdot)} \sup_{w_i(\cdot)}J_{x_i}^i(\rho_i, w_i, t),
\end{equation}
where $v_i(x_i, t)$ is the value (later on sporadically denoted also $v_i(t)$), and the minimization is performed over the Markovian controls for the reference player control problem, i.e. $\beta_{i_{\tau}}(\tau) = \rho_{i_{\tau}}(\tau) + w_{i_{\tau}}(\tau)$; here $\rho_i(\cdot)$ and $w_i(\cdot)$ are measurable functions from $R_0^+$ to $(R_0^+)^3$ which returns a transition rate and a penalty of the disturbance, respectively, at every time $t$.
\end{problem}
The Markov chain for the single player is defined as
\begin{equation} \label{eq:markovian}
\mathbb{P}[i_{\tau+h} = j|i_{\tau}] = [\rho_j(\tau)+w_j(\tau)]h +o(h).
\end{equation}
We define the generalised Legendre transform of the cost function $g(i, x_i, \rho_i)$ as
\begin{equation} \label{eq:h}
h(x_i, \Delta_iz, i) = \min_{\rho_i(\cdot)} g(i, x_i, \rho_i) + \rho_i^T \cdot \Delta_iz,
\end{equation}
where $\Delta_i: \mathbb{R}^3 \rightarrow \mathbb{R}^3$ is the difference operator on $i$ given by
\begin{equation} \label{eq:delta}
\Delta_iv = (v^1 - v^i, v^2 - v^i, v^3 - v^i)^T.
\end{equation}
When $v$ is the value function, we have that the Hamiltonian function $\mathcal{H}(\cdot)$ is given by:
\begin{equation} \label{eq:H}
\mathcal{H}(x_i, \Delta_iv, i) = \inf_{\rho_i(\cdot)} \sup_{w_i(\cdot)} g(\cdot) -\frac{1}{2}\|w_i\|^2_{\Gamma_i} + (\rho_i + w_i)^T \Delta_iv.
\end{equation}
Because of the superlinearity and uniform convexity of the cost function $g$, the function
\begin{equation} \label{eq:rho*i}
\rho^*_i(x_i, \Delta_iv, i) = \underset{\rho_i(\cdot)}{\operatorname{argmin}} \sup_{w_i(\cdot)} \quad g(\cdot) -\frac{1}{2}\|w_i\|^2_{\Gamma_i} + (\rho_i + w_i)^T \Delta_iv,
\end{equation}
is well defined except for its $i$th coordinate, which we denote by $\rho_{ii}$ and for which we assume:
\begin{equation} \label{eq:rho*ij}
\rho^*_{ii}(x, \Delta_iv, i) = -\sum_{j\neq i} \rho^{*T}_{ij}(x_j, \Delta_jv, j).
\end{equation}
We can now introduce the Hamilton-Jacobi ODEs as follows
\begin{equation} \label{eq:HJ}
\left\{\begin{array}{lll}
-\dot{v}_i =  \mathcal{H}(x_i, \Delta_iv, t), \\
v_i(T) = \psi(i, x_i).\\
\end{array}\right.
\end{equation}
A system of coupled ODEs with a terminal condition like the one in (\ref{eq:HJ}) is referred to as terminal value problem. The next result establishes that the solution of (\ref{eq:HJ}) is the value function, in accordance with \cite{Gomes}.\\
\begin{theorem}\label{th1}
Assume $\mathbf{x}(t): [0, T] \rightarrow \mathcal{S}^3$ is given over the horizon. Assume that $v: \mathbb{I}^3 \times [0, T] \rightarrow \mathbb{R}$ is a solution of the terminal value problem in (\ref{eq:HJ}). Then $v$ is the value function associated to the distribution $\mathbf{x}$, and the optimal Markovian control is $\beta_i^* = \rho_i^*+w_i^*$:
\begin{equation} \label{eq:rho*}
\begin{array}{lll}
\rho_i^* = - R_i^{-1} \Big[ \Delta_iv \Big]^- = 
\begin{scriptsize}
-\left[ \begin{array}{c}
R_{i1}^{-1} (v_1 - v_i)^- \\
R_{i2}^{-1} (v_2 - v_i)^- \\
R_{i3}^{-1} (v_3 - v_i)^- \end{array} \right],
\end{scriptsize}
\end{array}
\end{equation}
\begin{equation} \label{eq:w*}
\begin{array}{lll}
w_i^* =  \Gamma_i^{-1} \Big[ \Delta_iv \Big]^+ =
\begin{scriptsize}
\left[ \begin{array}{c}
\Gamma_{i1}^{-1} (v_1 - v_i)^+ \\
\Gamma_{i2}^{-1} (v_2 - v_i)^+ \\
\Gamma_{i3}^{-1} (v_3 - v_i)^+ \end{array} \right].
\end{scriptsize}
\end{array}
\end{equation}
\end{theorem}

\section{Stationary Mean-field Equilibrium}\label{sec:mfg}
\subsection{Mean-field Game}
When background players use strategy $\rho$ and the best response for the reference player is also $\rho$, we say that the current solution is a mean-field Nash equilibrium. The corresponding mean-field Nash equilibrium is given by the following system combining Kolmogorov and Hamilton-Jacobi equations:
\begin{equation} \label{eq:KHJ}
\left\{\begin{array}{lll}
\dot{x}_i(t) = (1 - x_1 - x_2)\beta_{3i} - x_i\beta_{i3}, \forall i \in \mathbb{I}^3\\
-\dot{v}_i(t) =  \mathcal{H}(x_i, \Delta_iv, t), \forall i \in \mathbb{I}^3\\
\mathbf{x}(0) = \mathbf{x}_0,\\
v_{i_T}(T) = \psi(i_T, x_{i_T}).\\
\end{array}\right.
\end{equation}
The above system is obtained from bringing together (\ref{eq:um2}) and (\ref{eq:rho*i}). Equation (\ref{eq:rho*i}) models the way in which individual players respond to a population behaviour described by (\ref{eq:um2}), and (\ref{eq:um2}) describes the way in which the population evolves as a whole under the assumption that all the players act according to (\ref{eq:rho*i}). The macroscopic dynamics are modelled by (\ref{eq:um2}), while the microscopic dynamics by (\ref{eq:rho*i}).
This problem is called initial-terminal value problem (ITVP) for the mean-field game. Expanding the Hamiltonian according to (\ref{eq:H}) and using the optimal control and disturbance from (\ref{eq:rho*})-(\ref{eq:w*}), the calculation of the value function is the following:
\begin{equation} \label{eq:vf}
\begin{array}{lll}
-\dot{v}_i =  \frac{1}{2}\|\rho_i^*\|^2_{R_i} - \frac{1}{2}\|w_i^*\|^2_{\Gamma_i} + (\rho_i^* + w_i^*)^T \Delta_iv + f_i(x_i)\\
\qquad =  -\frac{1}{2}\Big[R_{ij}^{-1}[(v_j - v_i)^-]^2 + R_{ik}^{-1}[(v_k - v_i)^-]^2 \Big] + \\
\qquad + \frac{1}{2}\Big[\Gamma_{ij}^{-1}[(v_j - v_i)^-]^2 + \Gamma_{ik}^{-1}[(v_k - v_i)^-]^2 \Big] + f_i(x_i) \\
\qquad = - \frac{1}{2}  (\Delta_iv)^{-^T} R_i^{-1}(\Delta_iv)^- + \frac{1}{2}  (\Delta_iv)^{+^T} \Gamma_i^{-1}(\Delta_iv)^+\\
\qquad+ f_i(x_i).
\end{array}
\end{equation}
Assume that $v_1, v_2<v_3$ and $R_{12}, R_{21}, \Gamma_{12}, \Gamma_{21} > 0$ sufficiently large, meaning that the cost to transition from 1 to 2 and vice versa is large. The specialised value function for each $i$ is:
\begin{equation} \label{eq:vfi}
\begin{array}{lll}
-\dot{v}_1 =  \frac{1}{2}\Gamma_{13}^{-1}(v_3 - v_1)^2 +f_1(x_1),\\
-\dot{v}_2 =  \frac{1}{2}\Gamma_{23}^{-1}(v_3 - v_2)^2 +f_2(x_2),\\
-\dot{v}_3 =  -\frac{1}{2}[R_{31}^{-1}(v_1 - v_3)^2 + R_{32}^{-1}(v_2 - v_3)^2] +f_3(x_3).\\
\end{array}
\end{equation}

By subtracting the third equation from the first and from the second in the above system we get:
\begin{equation} \label{eq:vij}
\left\{\begin{array}{lll}
\dot{v}_1-\dot{v}_3 = -\frac{1}{2}\Big{[}\Gamma_{13}^{-1}(v_3-v_1)^2 + R_{31}^{-1}(v_1-v_3)^2  \\
\qquad \qquad + R_{32}^{-1}(v_2-v_3)^2\Big{]} + f_3(x_3) - f_1(x_1),\\
\dot{v}_2-\dot{v}_3 = -\frac{1}{2}\Big{[}\Gamma_{23}^{-1}(v_3-v_2)^2 + R_{31}^{-1}(v_1-v_3)^2  \\ 
\qquad \qquad + R_{32}^{-1}(v_2-v_3)^2\Big{]} + f_3(x_3) - f_2(x_2).\\
\end{array}\right.
\end{equation}
Let us now denote by $y_1=v_1 - v_3$ and $y_2=v_2 - v_3$. System (\ref{eq:vij}) can then be rewritten as 
\begin{equation} \label{eq:yi}
\left\{\begin{array}{lll}
\dot{y}_1 =  -\frac{1}{2}a_{11}y_1^2 - \frac{1}{2}a_{12}y_2^2+c_1,\\
\dot{y}_2 =  -\frac{1}{2}a_{21}y_1^2 - \frac{1}{2}a_{22}y_2^2+c_2,\\
\end{array}\right.
\end{equation}
where $a_{11} = \Gamma_{13}^{-1}+R_{31}^{-1}$, $a_{12} = R_{32}^{-1}$, $a_{21} = R_{31}^{-1}$, $a_{22} = \Gamma_{23}^{-1}+R_{32}^{-1}$, $c_1 = f_3(x_3) - f_1(x_1)$ and $c_2 = f_3(x_3) - f_2(x_2)$.\\
System (\ref{eq:yi}) is used later on in Theorem \ref{th2} to establish existence of a stationary mean-field equilibrium. 

\subsection{Stationary solutions}
We are now interested in stationary solutions, also called stationary mean-field equilibrium points. These are defined as
\begin{equation} \label{eq:st}
\left\{\begin{array}{lll}
\sum_k x_k\beta_{ki}^* - \sum_j x_i\beta_{ij}^* = 0, \forall i \in \mathbb{I}^3, \\
\mathcal{H}(\mathbf{x}, \Delta_iv, t) = \kappa, \\
\end{array}\right.
\end{equation}
where $\kappa$ is a constant.
Note that, in a stationary mean-field equilibrium, function $c_i$ are fixed and constant as $\mathbf{x}$ is at an equilibrium of (\ref{eq:um2}) and therefore it is constant. As regards solutions of the first of (\ref{eq:st}), namely stationary distributions $\mathbf{x}(t)$, we have studied existence and stability in \cite{SB17M}. In the rest of this section we focus on the solution of the second of (\ref{eq:st}). We are ready to establish existence of stationary value functions, namely  value functions that satisfy  the second of (\ref{eq:st}). \\

\begin{theorem}\label{th2}
Let a stationary  distribution be given $t \mapsto \mathbf{x}(t) := \mathbf{\hat{x}} \in \mathcal{S}^3$. Then a stationary value function exists and is given by the mean-field game described by equations (\ref{eq:yi}) for the stationary problem (\ref{eq:st}):
\begin{equation} \label{eq:eqy}
 y_1^* = -\sqrt{\xi} =  -\sqrt{\Gamma_{23}^{-1}\Gamma_{13}y_2^{*^2} - 2\Gamma_{13}(f_2(x_2)-f_1(x_1))}.
\end{equation}
\end{theorem}
\begin{figure}[h]
\includegraphics[width=0.5\textwidth]{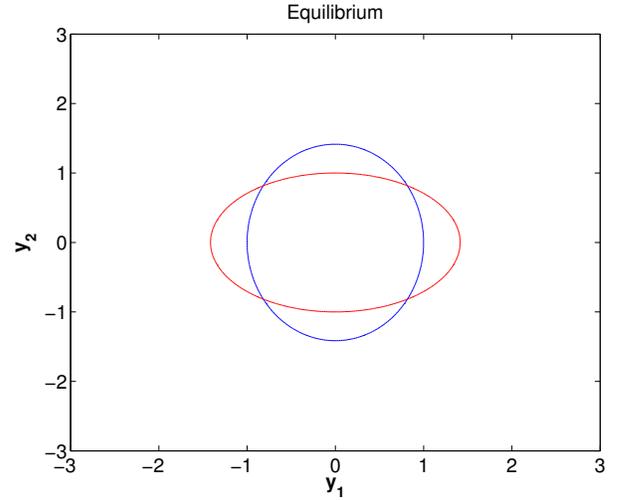}
\caption{Representation of the system of equations (\ref{eq:yi}).}
\label{fig:eq}
\end{figure}

We seek the equilibrium points at the intersections of two ellipses. Among the four points, we are interested in the one in the third quadrant. When $\Gamma_{13}^{-1}$ increases, the blue ellipse becomes thinner vertically, and similarly, when $\Gamma_{23}^{-1}$ increases, the red one becomes thinner horizontally. So, this happens when the inverse of both, i.e. $\Gamma_{13}$ and $\Gamma_{23}$ respectively, decreases. When the cost of the disturbance is lower, the equilibrium point shifts towards the axes, $y_2$ and $y_1$ respectively. If we view the difference of value functions between two nodes as  a potential difference, the potential difference is constant, i.e. the potential at the nodes is the same. It is easier for the disturbance to make the player go back to the node where their gain is lower. Further, $c_1$ and $c_2$ affect the distance of the ellipses from the center of axis, the equivalent of the radius in the circle. This is depicted in Fig. \ref{fig:eq}.
\subsection{Convergence to stationary solution}
In this section we study the link between the stationary and non stationary solutions. The next result, shows that the stationary solution is obtained from the nonstationary one in the limit.
Given that the monotonicity and contractivity assumptions hold in our case then we are in the presence of a \textit{contractive mean-field game}, see \cite{Gomes}. We can also say that the solution is unique (up to the addition of a constant to $v$).

\begin{theorem}\label{th3}
Given $T>0$, a vector $\mathbf{x}_0$ and a terminal condition $\psi$, let $(\mathbf{x}^T, v^T)$ be the solution of (\ref{eq:KHJ}) with initial-terminal conditions $\mathbf{x}^T(-T) = \mathbf{x}_0$ and $v^T_i(T) = \psi(i, x^T_i(T))$.
When $T\rightarrow \infty$
\begin{equation} \label{eq:itcu}
x^T(0)\rightarrow\bar{x} \qquad \|v^T(0) - \bar{v} \|_{\#}\rightarrow0,
\end{equation}
where $(\bar{x},\bar{v})$ is a solution of the stationary problem (\ref{eq:st}), and $\|\mathit{v}\|_{\#} = \inf_{\lambda \in \mathbb{R}} \|\mathit{v} +\lambda\mathbf{1}\|$.
The mean-field game is contractive, i.e. the stationary solution  (\ref{eq:st}) can be obtained when we study the asymptotic behaviour of the non-stationary solution of (\ref{eq:KHJ}).
\end{theorem}

\section{Honeybee Swarm}\label{sec:hs}
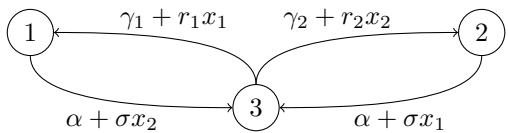
\begin{figure}[h]
\begin{centering}
      \begin{tikzpicture}
      \path 	(0,0) node[circle,draw](x) {$3$}
      		(-3,1) node[circle,draw] (z) {$1$}
		(3,1) node[circle,draw](y) {$2$};
	\draw[->,black] (x) .. controls +(up:1cm) and +(right:1cm) .. node[above] {$\gamma_1 + r_1x_1$} (z);
	\draw[->,black] (z) .. controls +(down:1cm) and +(left:1cm) .. node[below] {$\alpha + \sigma x_2$} (x);
	\draw[->,black] (x) .. controls +(up:1cm) and +(left:1cm) .. node[above] {$\gamma_2 +r_2x_2$} (y);
	\draw[->,black] (y) .. controls +(down:1cm) and +(right:1cm) .. node[below] {$\alpha + \sigma x_1$} (x);
      \end{tikzpicture}
\caption{Markov chain corresponding to the evolutionary model in \cite{SB17IFAC}.}
\label{markov2}
\end{centering}
\end{figure}

In this section, we link the more general model (\ref{eq:um}) to the collective decision-making model of honeybee swarm, as in \cite{SB17IFAC}, by showing that the latter can be a specific case of (\ref{eq:um}), when the parameters are appropriately chosen. We consider a population of honeybees and two options where to build a nest. Each option has a value. Scouting bees committed to one of the two options try to recruit the other bees, by using the so-called ``waggle dance''. To attract the uncommitted bees of the swarm, a cross-inhibitory signal is used by scouting bees committed to one option to disrupt the waggle dance of the bees committed to the other option. Bees in the uncommitted state can choose spontaneously to move to one of the two options. On the other hand, bees committed to one of the two options can spontaneously transition to the uncommitted state. We model the two options as two states, i.e. $1$ and $2$. The phenomena of the waggle dance and cross-inhibitory signals can be captured by using parameters $r := r_1, r_2$ and $\sigma := \sigma_1, \sigma_2$, respectively. Moreover, we model the spontaneous commitment by transition rates $\gamma_1, \gamma_2$ and the spontaneous act of transitioning to the uncommitted state by the transition rates $\alpha := \alpha_1, \alpha_2$. The Markov chain corresponding to the above model is depicted in Fig. \ref{markov2}.

Let us assume that the equilibrium point calculated in (\ref{eq:eqy}) involves the transition rates from the uncommitted state to state $1$ and state $2$. Then we have:
\begin{equation} \label{eq:y*sa}
y^*=\left[ \begin{array}{c}
y_1^*  \\
y_2^* \end{array} \right] =
\left[ \begin{array}{c}
v_1 - v_3  \\
v_2 - v_3 \end{array} \right] =
\left[ \begin{array}{c}
-\gamma_1-rx_1  \\
-\gamma_2-rx_2 \end{array} \right].
\end{equation}
The above means that the difference in the value functions between the uncommitted state and state $1$ and $2$ coincides with the transition rates from $3$ to $1$ and $2$, respectively. Furthermore, when $R_{31},R_{32}=1$, from (\ref{eq:rho*}), we have
\begin{equation} \label{eq:rho*sa}
\rho_3^*= - \left[ \begin{array}{c}
v_1 - v_3  \\
v_2 - v_3 \\
0\end{array} \right] =
\left[ \begin{array}{c}
\gamma_1+rx_1  \\
\gamma_2+rx_2  \\
0 \end{array} \right].
\end{equation}
Likewise, from (\ref{eq:w*}), we get:
\begin{equation} \label{eq:w*sa}
\begin{array}{lll}
w_1^*=\left[ \begin{array}{c}
0 \\
0 \\
\Gamma_{13}^{-1}(v_3 - v_1)^+ \end{array} \right] :=
\left[ \begin{array}{c}
0 \\
0 \\
\frac{\alpha + \sigma x_2}{\gamma_1+rx_1} (\gamma_1+rx_1) \end{array} \right] = \\
\qquad \qquad \qquad = \left[ \begin{array}{c}
0 \\
0 \\
\alpha + \sigma x_2 \end{array} \right],
\end{array}
\end{equation}
where the second equality is obtained by setting 
\begin{equation} \label{eq:Gamma}
\Gamma_{13} = (\gamma_1+rx_1)/(\alpha + \sigma x_2).
\end{equation}
Thus, the mean-field game model introduced in Section \ref{sec:mfm} yields the evolutionary model for the collective decision-making problem of the honeybee swarms, also studied in \cite{SB17ARXIV}, which we rewrite as 
\begin{equation} \label{eq:arxiv}
\left\{\begin{array}{lll}
\dot{x}_1 = x_3(x_1\rho_1 + \gamma_1) - x_1(x_2\sigma_2 + \alpha_1), \\
\dot{x}_2 = x_3(x_2\rho_2 + \gamma_2) - x_2(x_1\sigma_1 + \alpha_2), \\
\dot{x}_3 = x_1(x_2\sigma_2 + \alpha_1) + x_2(x_1\sigma_1 + \alpha_2) \\ \qquad -x_3(x_1\rho_1 + \gamma_1) -x_3(x_2\rho_2 + \gamma_2). \\
\end{array}\right.
\end{equation}
The above analogy also shows that the cross-inhibitory signal takes the form of a disturbance $w$ in the robust mean-field game  introduced in Section \ref{sec:mfm}. The analogy with the evolutionary model in \cite{Marshall} is exemplified by the Markov chain depicted in Fig. \ref{markov2}.\\
In the next subsection, we derive a network model with a finite number of players that approximate system (\ref{eq:um}), of which the mean-field game model in (\ref{eq:arxiv}) can be seen the asymptotic approximation when the number of players goes to infinity.

\subsection{Network Honeybee Swarm Model}
Let us consider system (\ref{eq:arxiv}), in the case where we have a finite number of players and let us now introduce the microscopic dynamics in the form of a network honeybee swarm model.\\
More specifically, the network model involves a population of $N$ players which correspond to the nodes of the network. An edge between two nodes indicates that the corresponding players interact with each others. The interaction network is described by a fully connected undirected graph $G$ with adjacency matrix $A$. The weights on each edge define the frequency of contact. In the following, let us denote $r_i$, $s_i$ and $z_i$ the probability that player $i$ is in state $1$, $2$ and $3$, respectively. The transitions rates to a committed and uncommitted state are modelled by parameter $\beta_{ij}$, and represent the transition rates from state $i$ to state $j$ as presented in Section~\ref{sec:mfm}.\\
The model describing the time evolution of the infection probability for each node is given by the following system of equations:
\begin{equation} \label{eq:CDM}
\left\{\begin{array}{lll}
\dot{s}_i(t) = -\beta_{23}' s_i(t) \sum_{j=1}^N a_{ij}r_j(t) +\beta_{32}' z_i(t) \sum_{j=1}^N a_{ij}s_j(t) \\
\qquad \quad -\beta_{23}'' s_i(t) + \beta_{32}'' z_i(t), \\
\dot{z}_i(t) =  \beta_{23}' s_i(t) \sum_{j=1}^N a_{ij}r_j(t) -\beta_{32}' z_i(t) \sum_{j=1}^N a_{ij}s_j(t) \\
\qquad \quad +\beta_{13}' r_i(t) \sum_{j=1}^N a_{ij}s_j(t) -\beta_{31}' z_i(t) \sum_{j=1}^N a_{ij}r_j(t) \\
\qquad \quad +\beta_{23}'' s_i(t) - \beta_{32}'' z_i(t) +\beta_{13}'' r_i(t) - \beta_{31}'' z_i(t), \\
\dot{r}_i(t) = -\beta_{13}' r_i(t) \sum_{j=1}^N a_{ij}s_j(t) +\beta_{31}' z_i(t) \sum_{j=1}^N a_{ij}r_j(t) \\
\qquad \quad -\beta_{13}'' r_i(t) + \beta_{31}'' z_i(t).\\
\end{array}\right.
\end{equation}
In the equation for $\dot s_i$, the first term accounts for the cross-inhibitory signal, the second term describes the waggle dance and the last two terms accounts for the spontaneous abandon of the commitment and the spontaneous commitment itself, respectively. Similar comments apply to the equation for $\dot r_i$ and $\dot z_i$. We can rewrite system (\ref{eq:CDM}) in vector form as
\begin{equation} \label{eq:CDMvf1}
\left\{\begin{array}{lll}
\dot{s}(t) = -\beta_{23}' diag(s(t))Ar(t) + \beta_{32}' diag(z(t))As(t)\\
\qquad \quad -\beta_{23}'' s(t) + \beta_{32}'' z(t), \\
\dot{z}_i(t) =  +\beta_{23}' diag(s(t)) Ar(t) -\beta_{32}' diag(z(t)) As(t) \\
\qquad \quad +\beta_{13}' diag(r(t)) As(t) -\beta_{31}' diag(z(t)) Ar(t) \\
\qquad \quad +\beta_{23}'' s(t) - \beta_{32}'' z(t) +\beta_{13}'' r(t) - \beta_{31}'' z(t), \\
\dot{r}(t) = -\beta_{13}' diag(r(t))As(t) + \beta_{31}' diag(z(t))Ar(t)\\
\qquad \quad -\beta_{13}'' r(t) + \beta_{31}'' z(t). \\
\end{array}\right.
\end{equation}
We can now present some results on the impact of the interaction topology to the collective decision-making process.
\begin{theorem}\label{th4}
Let us consider the network honeybee swarm model (\ref{eq:CDMvf1}), given $\beta_{ij}', \beta_{ij}'' > 0$ for all $i$, $j$, over a strongly connected graph with adjacency matrix $A$. 
The following statements hold:
\begin{enumerate}
\item If $s(0), r(0) \in [0,1]_n$, then $s(t), r(t) \in [0,1]_n$ for all $t>0$, where $[0,1]_n$ is the $n-$dimensional space $[0,1]$. 
\item The set of equilibrium points is the set of pairs $(\bar 1_n, \bar 0_n)$, $(\bar 0_n, \bar 1_n)$. The equilibrium points are asymptotically stable.
\item Let $\beta_{23}' = k \beta_{32}'$ and $\beta_{23}'' = k \beta_{32}''$, where $k$ is a parameter that corresponds to a measure of the connectivity. The set of equilibrium points includes the set of pairs $((1/(2+k))_n, (1/(2+k))_n)$. These equilibrium points are asymptotically stable.
\end{enumerate}
\end{theorem}
The novelty of the above result is that a higher connectivity determines higher values of $z$ at the equilibrium. Thus, the higher the connectivity, the larger the probability of a generic agent to be in the uncommitted state. 

\section{Virus Propagation}\label{sec:vp}
Here we introduce the virus propagation model under a network topology, which can be obtained from the model presented in Section~\ref{sec:mfm} by assuming that the cross-inhibitory signal is negligible, i.e. all $\beta'_{ij}$ coefficients are zero. The virus propagation models we refer to are the \emph{SIS} and the \emph{SIR} models, and their variants, see \cite{FBLN}. In the \emph{SIS} model, the individuals can either be susceptible, or infected. The two states are denoted by $s$ and $z$, respectively. After being infected by the virus, $n$ infected individuals can return to the susceptible state. The \emph{SIR} model describes the case where the individuals who recover from the virus cannot be infected again. 
Let us assume that the equilibrium point calculated in (\ref{eq:eqy}) is given by:
\begin{equation} \label{eq:y*sa2}
y^*=\left[ \begin{array}{c}
y_1^*  \\
y_2^* \end{array} \right] =
\left[ \begin{array}{c}
v_1 - v_3  \\
v_2 - v_3 \end{array} \right] =
\left[ \begin{array}{c}
-\beta_{31}  \\
-\beta_{32} \end{array} \right].
\end{equation}
Assuming that $R_{31},R_{32}=1$, from (\ref{eq:rho*}) it can be seen that
\begin{equation} \label{eq:rho*sa2}
\rho_3^*= - \left[ \begin{array}{c}
v_1 - v_3  \\
v_2 - v_3 \\
0\end{array} \right] =
\left[ \begin{array}{c}
\beta_{31}  \\
\beta_{32} \\
0 \end{array} \right].
\end{equation}
Likewise, from (\ref{eq:w*}), we get:
\begin{equation} \label{eq:w*sa2}
\begin{array}{lll}
w_1^*=\left[ \begin{array}{c}
0 \\
0 \\
\Gamma_{13}^{-1}(v_3 - v_1)^+ \end{array} \right] :=
\left[ \begin{array}{c}
0 \\
0 \\
\frac{\beta_{13}x_3}{\beta_{31}} \beta_{31} \end{array} \right] = \\
\qquad \qquad \qquad = \left[ \begin{array}{c}
0 \\
0 \\
\beta_{13}x_3 \end{array} \right].
\end{array}
\end{equation}
In the above  we have set 
\begin{equation} \label{eq:Gamma2}
\Gamma_{13} = \frac{\beta_{13}x_3}{\beta_{31}}.
\end{equation}
The resulting model is as the one in (\ref{eq:um}). The infection rate can thus be linked to the matrix $\Gamma$ which is the cost matrix for the adversarial disturbance.
In the next subsection, we will tackle the problem of virus propagation in a network and introduce the microscopic dynamics describing the time evolution of the probabilities of an individual to be in one of the three aforementioned states.

\subsection{Network Cyber-Attack Model}
Let us review the model of virus propagation in the context of cyber-security. We model the problem of a hacker trying to disrupt two \emph{holons}. A \emph{holon} is an intelligent entity which is seen as an autonomous unit component of a multi-component system. It is autonomous due to the fact that it is able to process information and make decision on its own. In our context, a holon is part of a power grid or information system, which is also autonomous. 
Variables $r$, $s$ and $z$ represent the probability of an individual of being in state $1$, $2$ and $3$, respectively. Parameters $\beta_{31}, \beta_{32}$ models the capability of the holons to recover from the infection or to mitigate the cyber-attacks. Here, we model a cyber-attack at two holons, with the aim of disrupting the services or the data stored. The final values of $s, r$ determine how far the corruption of data has propagated through the network. We consider two types of attacks which can be modelled by different values of $\beta_{ij}$, see \cite{RP17}:
\begin{itemize}
\item \textbf{sequential attack}: the hacker sends continuous burst attacks, disrupting the customer layer and the information of control signals. The two holons under attack are characterized by high values of $\beta_{13}, \beta_{23}$ for short periods of time;
\item \textbf{continuous low rate attack}: usually aimed at disrupting voltage control at the holons; here small values of $\beta_{13}, \beta_{23}$ endures for the entire time horizon.
\end{itemize}

Let (\ref{eq:y*sa2}) hold. Under the assumption that the cross-inhibitory signal is negligible we henceforth explore the role of the connectivity in the equilibrium and stability properties of the system. Under no cross-inhibitory signal, system (\ref{eq:CDMvf1}) reduces to
the following system of ODEs:
\begin{equation} \label{eq:VPN1}
\left\{\begin{array}{lll}
\dot{s}_i(t) = -\beta_{23} s_i(t) \sum_{j=1}^N a_{ij}z_j(t) + \beta_{32} z_i(t), \\
\dot{z}_i(t) =  \beta_{23} s_i(t) \sum_{j=1}^N a_{ij}z_j(t) + \beta_{13} r_i(t) \sum_{j=1}^N a_{ij}z_j(t)\\ \qquad \quad - \beta_{32} z_i(t) - \beta_{31} z_i(t), \\
\dot{r}_i(t) = - \beta_{13} r_i(t) \sum_{j=1}^N a_{ij}z_j(t) + \beta_{31} z_i(t),\\
\end{array}\right.
\end{equation}
where $s_i, r_i$ are the variables describing the holon. By considering the conservation of mass, the above system can be rewritten in vector form as
\begin{equation} \label{eq:VPNvf1}
\left\{\begin{array}{lll}
\dot{s}(t) = -\beta_{23}diag(s(t))Az(t) + \beta_{32} z(t), \\
\dot{r}(t) = -\beta_{13}diag(r(t))Az(t) + \beta_{31} z(t), \\
\end{array}\right.
\end{equation}
where $z(t) = 1 - s(t) - r(t)$.

\begin{theorem}\label{th5}
Let us consider the virus propagation network model (\ref{eq:VPNvf1}), with $\beta_{ij} > 0$ for all $i$, $j$, over a strongly connected graph with adjacency matrix $A$. 
The following statements hold:
\begin{enumerate}
\item If $s(0), r(0) \in [0,1]_n$, then $s(t), r(t) \in [0,1]_n$ for all $t>0$. 
\item The set of equilibrium points is the set of pairs $(s^*, \bar 1_n - s^*)$, for any $s^* \in [0,1]_n$, and the linearization of (\ref{eq:VPNvf1}) about $(s^*,  \bar 1_n - s^*)$ is
\begin{equation} \label{eq:VPNvf2}
\left\{\begin{array}{lll}
\dot{s}(t) = -\beta_{23}diag(s^*)Az(t) + \beta_{32} z(t), \\
\dot{r}(t) = -\beta_{13}diag(\bar 1_n - s^*)Az(t) + \beta_{31} z(t). \\
\end{array}\right.
\end{equation}
For the complete system (\ref{eq:VPN1}), the set of equilibrium points is the set of tuples $(s^*, \bar 0_n, \bar 1_n - s^*)$. Furthermore, the above equilibrium points are asymptotically stable if the following condition holds:
\begin{equation} \label{eq:VPNc}
\begin{array}{lll}
\beta_{23}diag(s^*)A \bar 1_n + \beta_{13}diag(\bar 1_n - s^*)A \bar 1_n \\ \qquad \qquad \qquad < (\beta_{32} + \beta_{31}) \bar 1_n.
\end{array}
\end{equation}
\item If $\lim_{\beta_{32},\beta_{13} \rightarrow 0}$, then system (\ref{eq:VPN1}) can be approximated by the standard $SIR$ model, see \cite{FBLN}.
\end{enumerate}
\end{theorem}
To get a physical understanding of the previous theorem, consider inequality (\ref{eq:VPNc}). When the connectivity is higher, which in turn implies higher values of the entries of the adjacency matrix, the left-hand side of the inequality increases, eventually becoming greater than the right-hand side. This means that the system becomes unstable, since the higher the connectivity, the higher the probability that a cyber-attack can infect a holon in the network.

\section{Numerical Analysis}\label{NA}
In this section, we carry out two sets of simulations of the virus propagation scenario. More specifically, in the first set we study the continuous low-rate attacks while in the second set we focus on the sequential attacks.

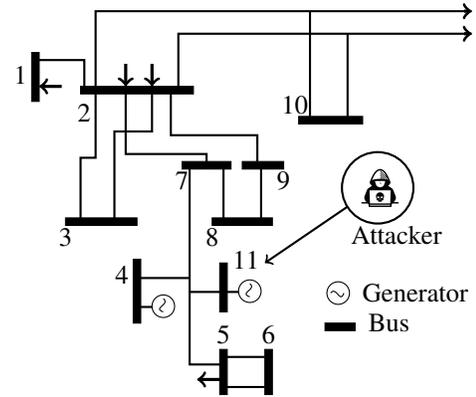
\begin{figure} [h]
\centering
\begin{tikzpicture}
\begin{scope}[shift={(0,0)},scale=.5]
\draw[fill=black] (-3.6,-0.5) rectangle (-2.5,-.3);
\node at (-2.5,-.8){9}; \draw [thick]  (-3.2,-.3) -- (-3.2,.4) -- (-5.5,0.4) -- (-5.5,1.5);
\begin{scope}[shift={(-4.2,-3.6)},scale=1,rotate=-90]
\draw (.15,0.8) circle (.3cm); \draw [thick]  (.17,.5) -- (.17,0);
\draw (-.05,0.8) .. controls (.1,1.0) and (.2,.5) .. (.3,0.8);
\draw[fill=black] (-.6,0) rectangle (0.7,.2);
\node at (-0.7,0.7){11};\draw [thick]  (0.17,.0) -- (0.17,-0.8);
\end{scope}
\draw [thick] (0,-1) circle (.95cm); \draw [thick, ->]  (-.8,-1.5) -- (-3,-3);
\node[inner sep=0pt] (russell) at (0,-1){\includegraphics[width=.035\textwidth]{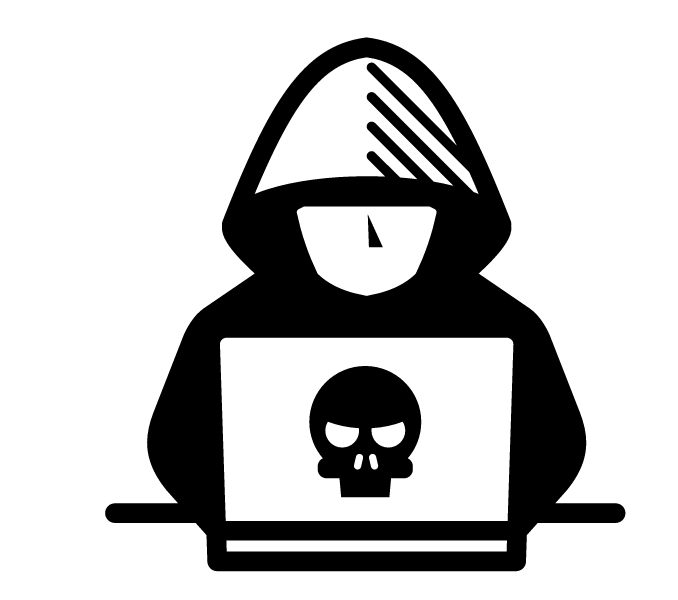}};
\node at (.5,-2.25){Attacker};
\begin{scope}[shift={(-4.2,-5.2)},scale=1,rotate=-90]
\draw[fill=black] (.1,0) rectangle (1.3,.2);\draw [very thick, ->]  (.9,0) -- (.9,-0.6);
\node at (-.3,0.1){5}; \draw [thick]  (1.1,.0) -- (1.1,1.2);
\end{scope}
\begin{scope}[shift={(-3,-5.2)},scale=1,rotate=-90]
\draw[fill=black] (.1,0) rectangle (1.3,.2);
\node at (-.3,0.1){6}; \draw [thick]  (0.3,.0) -- (0.3,-1.2);
\end{scope}
\begin{scope}[shift={(-6.5,-4)},scale=1,rotate=-90]
\draw (.15,0.8) circle (.3cm); \draw [thick]  (.17,.5) -- (.17,0);
\draw (-.05,0.8) .. controls (.1,1.0) and (.2,.5) .. (.3,0.8);
\draw[fill=black] (-1.1,0) rectangle (0.6,.2);
\node at (-0.7,-.3){4}; \draw [thick]  (-0.6,.0) -- (-0.6,1.5) ;
\end{scope}
\begin{scope}[shift={(-3.7,-2)},scale=1,rotate=0]
\draw[fill=black] (-.7,0) rectangle (0.9,.2);
\node at (-0.7,-.3){8};\draw [thick]  (-0.4,.0) -- (-0.4,1.5); \draw [thick]  (0.6,.0) -- (0.6,1.5);
\end{scope}
\begin{scope}[shift={(-4.5,-.5)},scale=1,rotate=0]
\draw[fill=black] (-.7,0) rectangle (0.6,.2);
\node at (-0.7,-.3){7};\draw [thick]  (-0.05,.0) -- (-0.05,0.4) -- (-2.2,0.4) -- (-2.2,2.1); \draw [thick]  (-0.5,.0) -- (-0.5,-5.2) -- (0.4,-5.2);
\end{scope}
\begin{scope}[shift={(-8,-2.0)},scale=1,rotate=0]
\draw[fill=black] (-.3,0) rectangle (1.6,.2);
\node at (-0.3,-.3){3}; \draw [thick]  (1,.0) -- (1,2.5) -- (2,2.5) -- (2,3.5);
\end{scope}
\begin{scope}[shift={(-7,1.5)},scale=1,rotate=0]
\draw[fill=black] (-.9,0) rectangle (2.1,.2);
\draw [very thick, ->]  (.3,0.8) -- (.3,0.2);
\draw [very thick, ->]  (1,0.8) -- (1,0.2);
\node at (-.8,-.5){2};\draw [thick]  (-0.5,.0) -- (-0.5,-1.7) -- (-0.9,-1.7) -- (-0.9,-3.3); \draw [thick]  (-0.5,.0) -- (-0.5,2.2) -- (9.5,2.2); \draw [thick]  (1.7,.0) -- (1.7,1.6) -- (9.5,1.6);
\draw [very thick, ->]  (9.3,2.2) -- (9.5,2.2); \draw [very thick, ->]  (9.3,1.6) -- (9.5,1.6);
\end{scope}
\begin{scope}[shift={(-4,2.0)},scale=1,rotate=0]
\draw[fill=black] (1.9,-1.1) rectangle (3.6,-1.3);
\node at (1.8,-.8){10};\draw [thick]  (2.2,-1.1) -- (2.2,1.7); \draw [thick]  (3.2,-1.1) -- (3.2,1.1);
\end{scope}
\begin{scope}[shift={(-9,2.0)},scale=1,rotate=90]
\draw[fill=black] (-.7,0) rectangle (0.6,.2);\draw [very thick, ->]  (-.3,-0.6) -- (-.3,0);
\node at (0,0.5){1};\draw [thick]  (0.4,.0) -- (0.4,-1.2) -- (-0.3,-1.2);
\end{scope}
\begin{scope}[shift={(-1.2,-4.6)},scale=1,rotate=0]
\draw (.15,0.8) circle (.3cm); \draw (-.05,0.8) .. controls (.1,1.0) and (.2,.5) .. (.3,0.8);
\node at (2.2,0.8){Generator};
\end{scope}
\begin{scope}[shift={(-1.2,-4.8)},scale=1,rotate=0]
\draw[fill=black] (-0.2,0) rectangle (0.6,.2);
\node at (1.5,0.2){Bus};
\end{scope}
\end{scope}
\end{tikzpicture}
\caption{Line diagram representation of the Walpole GSP - Peterborough (EPN), from the Regional Development Plan in \cite[p 18]{RDP14}. Attacker image from \emph{https://thenounproject.com/term/hacker/870666/}.}  \label{fig:LD}
\end{figure}

\begin{figure} [h]
\centering
\begin{tikzpicture}
\begin{scope}[shift={(0,0)},scale=.5]
\draw[dotted, ultra thick] (-3,0) circle (5cm);
\begin{scope}[shift={(-3.5,-1.5)},scale=1,rotate=0]
\draw[fill=gray, ultra thick] (.15,0.8) circle (.3cm); \draw [ultra thick]  (-.1,.8) -- (-4.2,.35);  \draw [ultra thick]  (0,.55) -- (-1.5,-.3);
\node at (0,1.5){7}; 
\end{scope}
\begin{scope}[shift={(0,1)},scale=1,rotate=0]
\draw[fill=gray, ultra thick] (.15,0.8) circle (.3cm); \draw [ultra thick]  (0,.6) -- (-.8,-.8);\node at (0,1.5){9};
\end{scope}
\begin{scope}[shift={(-1.5,2.5)},scale=1,rotate=0]
\draw[fill=gray, ultra thick] (.15,0.8) circle (.3cm); \draw [ultra thick]  (-.1,.6) -- (-1.3,-1.1);
\node at (.7,1.2){10};
\end{scope}
\begin{scope}[shift={(-.5,-5)},scale=1,rotate=0]
\draw[fill=gray, ultra thick] (.15,0.8) circle (.3cm); \draw [ultra thick]  (0,1) -- (-2.7,4); \draw [ultra thick]  (-.15,.75) -- (-7.2,3.7); \draw [ultra thick]  (-.1,.85) -- (-4.5,3.2);
\node at (0.6,0.3){11};
\end{scope}
\begin{scope}[shift={(.5,-5)},scale=1,rotate=0]
\draw (.15,0.8) circle (.3cm); \draw [thick]  (-.6,0.8) -- (-.15,0.8);
\draw (-.05,0.8) .. controls (.1,1.0) and (.2,.5) .. (.3,0.8);
\end{scope}
\begin{scope}[shift={(-1,-.8)},scale=1,rotate=0]
\draw[fill=gray, ultra thick] (.15,0.8) circle (.3cm); \draw [ultra thick]  (-.1,.7) -- (-2.1,0.15); 
\node at (0,1.5){8};
\end{scope}
\begin{scope}[shift={(-3,0.5)},scale=1,rotate=0]
\draw[fill=gray, ultra thick] (.15,0.8) circle (.3cm); \draw [ultra thick]  (-.15,.8) -- (-2.7,0.3); \draw [ultra thick]  (.1,.5) -- (-.3,-0.9); \draw [ultra thick]  (.4,.8) -- (2.85,1.3); \draw [ultra thick]  (-.1,.95) -- (-1.85,2.2);
\node at (0,1.5){2};
\end{scope}
\begin{scope}[shift={(-6,0)},scale=1,rotate=0]
\draw[fill=gray, ultra thick] (.15,0.8) circle (.3cm); \draw [ultra thick]  (.4,.6) -- (2.4,-.6);
\node at (0,1.5){3};
\end{scope}
\begin{scope}[shift={(-8,-2.0)},scale=1,rotate=0]
\draw[fill=gray, ultra thick] (.15,0.8) circle (.3cm); 
\node at (-.3,1.3){4};
\end{scope}
\begin{scope}[shift={(-9,-2.0)},scale=1,rotate=0]
\draw (.15,0.8) circle (.3cm); \draw [thick]  (.45,0.8) -- (.8,0.8);
\draw (-.05,0.8) .. controls (.1,1.0) and (.2,.5) .. (.3,0.8);
\end{scope}
\begin{scope}[shift={(-5,-2.5)},scale=1,rotate=0]
\draw[fill=gray, ultra thick] (.15,0.8) circle (.3cm); \draw [ultra thick]  (.4,0.7) -- (.9,.6);\draw [ultra thick]  (-.1,0.9) -- (-2.6,1.25);
\node at (0,1.4){5};
\end{scope}
\begin{scope}[shift={(-4,-2.7)},scale=1,rotate=0]
\draw[fill=gray, ultra thick] (.15,0.8) circle (.3cm); 
\node at (.8,.7){6};
\end{scope}
\begin{scope}[shift={(-5,2.0)},scale=1,rotate=0]
\draw[fill=gray, ultra thick] (.15,0.8) circle (.3cm); 
\node at (0,1.5){1};
\end{scope}
\end{scope}
\end{tikzpicture}
\caption{Graph representation of part of the Walpole GSP - Peterborough (EPN), from the Regional Development Plan in \cite[p 18]{RDP14}.}  \label{fig:graphr}
\end{figure}
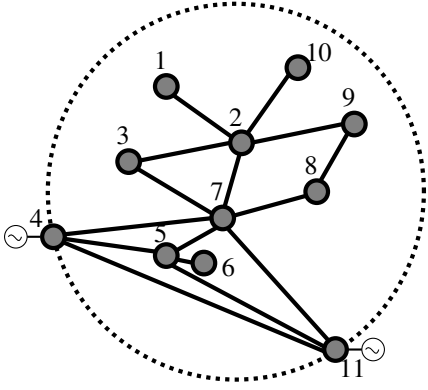

Let us consider the virus propagation model as in (\ref{eq:VPN1}). For this set of simulations, we use the network Walpole GSP - Peterborough (EPN), from the Regional Development Plan in \cite[p 18]{RDP14}. Figures \ref{fig:LD}-\ref{fig:graphr} show the line diagram and the graph representation of the network, respectively. 
\begin{table} [h]
\begin{center}
   \begin{tabular}{|c|c|c|c|}
   \hline
   Parameter & I & II \\
   \hline \hline
   $\beta_{13}$ & $0.13$ & $\{0.13, 0.65\}$\\ 
   \hline
   $\beta_{23}$ & $0.13$ & $\{0.13, 0.65\}$\\ 
   \hline
   \end{tabular}
\end{center}
\caption{Varying parameters for the simulations.}
\label{t:data1}
\end{table}

\begin{figure}[h]
\includegraphics[width=0.45\textwidth]{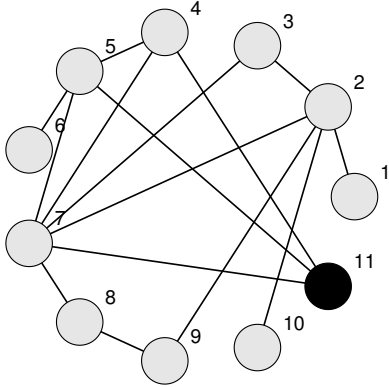}
\vspace{-.5cm}
\caption{Initial configuration of the network: the only infected node is node~$11$.}
\label{fig:n_init}
\end{figure}

\noindent \textbf{Network}. To show the infection probability at each node we color the nodes in greyscale. In particular, \emph{black} indicates that a node has a value of infection greater than $0.75$, \emph{dark grey} denotes a value of infection in the interval between $0.5$ and $0.75$, \emph{grey} indicates a value of infection in the interval between $0.25$ and $0.5$; finally, \emph{white} denotes a value of infection below $0.25$. The simulation plot of the initial network configuration is presented in Fig. \ref{fig:n_init}: as it can be seen from the plot, we assume that node $11$ is the infected node which is obtained by setting his state equal to $(0,1,0)$, the initial state of each of the first five nodes is $(0,0,1)$ and for the last five is $(1,0,0)$. For all the following scenarios, $\beta_{31}, \beta_{32}$ are fixed and set to $0.1$, to model the curing rates at the nodes. All the parameters varying across the simulations are shown in Table~\ref{t:data1}. In order to present the results in a consistent way, each network plot is supported by a histogram showing the number of nodes in each of the four categories explained above: no infection for a value below $0.25$ (white), small for a value between $0.25$ and $0.5$ (grey), moderate for a value between $0.5$ and $0.75$ (dark grey) and high for a value above $0.75$ (black).\\

\noindent \textbf{Frequency}. We present a set of simulations involving the frequency analysis of the buses in the network after the attacks. To study this phenomenon, we model the buses as coupled oscillators, as in \cite{FBLN}. The frequency of each bus is 50 hertz (Hz), with possible oscillations in the interval $[49.5, 50.5]$. Going above or lower these values can compromise the stability of the network, causing high damage to the provider and customers. In the original model, the frequency of each bus $\dot \theta_i$ evolves as in the following ODE:
\begin{equation} \label{eq:freq}
\begin{array}{lll}
\ddot \theta_i = \frac{\omega_i}{M_i} - \frac{K}{NM_i} \sum_{j=1}^N sin(\theta_i - \theta_j) - \frac{D_i}{M_i}\dot \theta_i,
\end{array}
\end{equation}
where $\omega_i$ is the natural frequency, $K$ is a parameter, $N$ is the number of nodes in the network, $M_i$ and $D_i$ are the inertia and damping coefficients. The quotient $K/N$ is the coupling strength and it is equal for all the nodes. In our case, we develop a model to include the adversarial disturbance caused by the cyber-attacks and the fact that we assume that all the buses have the same natural frequency.
\begin{equation} \label{eq:freq2}
\begin{array}{lll}
\ddot \theta_i = \frac{\omega}{M} - \frac{K}{NM} \sum_{j=1}^N sin(\theta_i - \theta_j) - \frac{D}{M}\dot \theta_i + \zeta_i,
\end{array}
\end{equation}
where $\zeta_i$ is the added disturbance. For each iteration, the input of this model is the output of (\ref{eq:VPN1}) at steady state in terms of the probability of infection for every node. The attack is modelled through a disturbance of the type
\begin{equation} \label{eq:noise}
\begin{array}{lll}
\zeta_i = \pm \hat k \hat \omega,
\end{array}
\end{equation}
which can be tuned to model the strength of the attack. In the simulations, the thick dash-dot blue line denotes the evolution of the frequency of node $1$ and the thick green line the evolution of node $7$, while the other thin dashed lines are for the other nodes. The two nodes have been chosen because the former has only one connection, while the latter is the most connected node in the network.\\

\begin{figure}[h!]
        \centering
         \begin{subfigure}[h]{0.45\textwidth}  
            \centering 
            \vspace{-0.8cm}
            \includegraphics[width=1.05\textwidth]{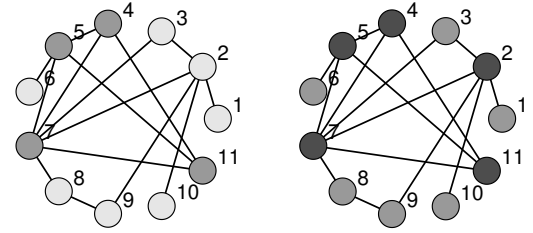}
            \vspace{-2cm}
            \caption[I small]%
            {{\small Continuous low rate attack: after several iterations (left) the infection spreads and at the end of the horizon (right) the infection is relatively contained.}}    
            \label{fig:n_clra}
        \end{subfigure}
        \begin{subfigure}[h]{0.45\textwidth}
            \centering
            \vspace{0.3cm}
            \includegraphics[width=\textwidth, height=6.25cm]{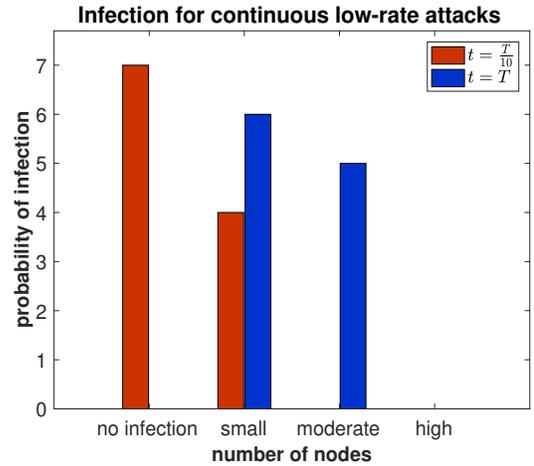}
            \caption[I big]%
            {{\small Histogram showing the infection percentage for the nodes in the network after several iterations and at the end of the horizon.}}    
            \label{fig:h_clra}
        \end{subfigure}
        \begin{subfigure}[h]{0.45\textwidth}
            \centering
            \vspace{0.3cm}
            \includegraphics[width=\textwidth]{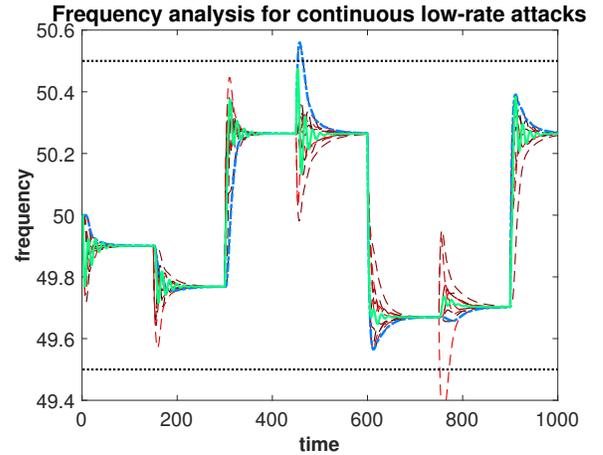}
            \caption[I big]%
            {{\small Frequency analysis: the impact of the attack, measured by the amplitude of the frequency of oscillations, is higher for node 1, characterized by lower connectivity (dash-dot line), than for node 7, characterized by higher connectivity (thick green line).}}    
            \label{fig:f_clra}
        \end{subfigure}
        \caption[ Continuous low-rate attacks: histogram and network. ]
        {\small Behaviour of the system when the attacks are of the type continuous low-rate.}
        \label{fig:clra}
\end{figure} 
\noindent \textbf{Continuous low rate attack}. As depicted in Fig. \ref{fig:n_clra}, small values of $\beta_{13}, \beta_{23}$ endure for the whole time horizon, more specifically we set $\beta_{13}, \beta_{23} = 0.1$. The infection spreads from node $11$, but the curing rates $\beta_{31}, \beta_{32}$ have values close to the infection rates, such that the probability of infection at nodes remains low. Note that the nodes with more links are those that are more likely to be infected. As it can be seen in the histogram, i.e. Fig. \ref{fig:h_clra}, after several iterations, only 4 nodes, i.e. nodes 4, 5, 7 and 11 from the network plot, have a small infection value, while the rest 7 have no infection at all; however, at the end of the horizon, the infection has spread and the nodes have either a small or a moderate infection value. For the frequency analysis, shown in Fig. \ref{fig:f_clra}, the probability of attack is sampled every 150 iterations. As it can be seen, even though the probability of infection of node $7$ is almost twice as much as the one of node $1$, and thus it is more likely to have an attack at every sample, the higher connectivity attenuates the strength of the attack.\\

\begin{figure}[h!]
        \centering
        \begin{subfigure}[h]{0.45\textwidth}  
            \centering
            \vspace{-1cm} 
            \includegraphics[width=1.05\textwidth]{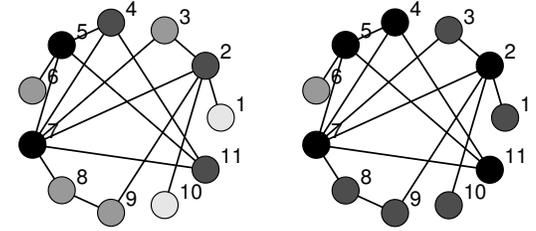}
            \vspace{-2cm}
            \caption[I small]%
            {{\small Sequential attack: after several iterations (left) the infection spreads and at the end of the horizon (right) the infection has affected most of the nodes with moderate or high probability.}}    
            \label{fig:n_sa}
        \end{subfigure}
        \begin{subfigure}[h]{0.45\textwidth}
            \centering
            \vspace{0.3cm}
            \includegraphics[width=\textwidth, height=6.25cm]{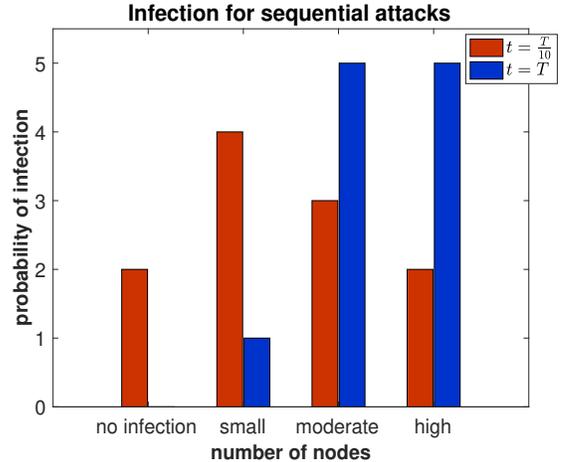}
            \caption[I big]%
            {{\small Histogram showing that, at $t=T$, the infection has affected most of the nodes with moderate or high probability.}}    
            \label{fig:h_sa}
        \end{subfigure}
        \begin{subfigure}[h]{0.45\textwidth}
            \centering
            \vspace{0.3cm}
            \includegraphics[width=\textwidth]{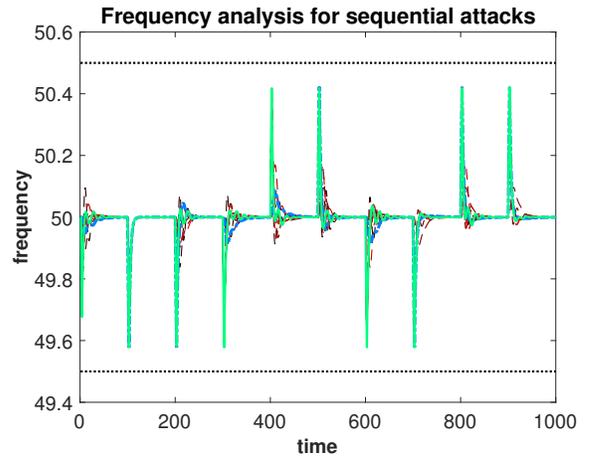}
            \caption[I big]%
            {{\small Frequency analysis: when node $7$ (thick green line) and other nodes are attacked, the first one has lower peaks compared to the others with lower connectivity.}}    
            \label{fig:f_sa}
        \end{subfigure}
        \caption[ Sequential attacks: histogram and network. ]
        {\small Behaviour of the system when the attacks are of the type sequential.}
        \label{fig:sa}
\end{figure}

\noindent \textbf{Sequential attack}. This set of simulations is shown in Fig.~\ref{fig:sa}. To model sequential burst attacks, we set $\beta_{13}, \beta_{23}$ five times greater than their normal value every five iterations. Consider the similarities with the most common denial of service (DOS) attacks aiming at disrupting the customer layer or the information of control signals for a service which becomes unavailable, i.e. a website where the server does not respond. In this scenario, as depicted in Fig.~\ref{fig:n_sa}, at the end of the time horizon, there are more nodes with higher probability of infection, the higher the connectivity the larger the infection values at the nodes. This can also be observed from the values of the histogram in Fig.~\ref{fig:h_sa}. In terms of the frequency, shown in Fig.~\ref{fig:f_sa}, for this type of attack the probability of infection is sampled every 100 iterations with higher value of $\hat k$. The connectivity plays an important role in this case, too: when node $7$ and other nodes are attacked, the first one has lower peaks, as it can be seen when comparing its peaks to those of node 1 around iterations 500 and 800.\\

\section{Conclusion}\label{sec:conc}
For a continuous-time dynamic game framework, we have presented a mean-field game model, we have introduced the \emph{initial-terminal value problem} and the corresponding stationary solution, with the aim of studying the stability. Under specific assumptions on the transition rates and on the cost matrices, we have linked the current model to existing models in the literature such as the bio-inspired model of honeybee swarm and to the cyber-security model for virus propagation. By introducing an underlying network, we studied the stability and provided numerical analysis for a case study of the network Walpole GSP - Peterborough (EPN). The plan for future works includes i) the analytical study of the region of convergence, and ii) the extension of the results to stochastic dynamics with births and deaths in the population, and iii) the study of the impact of the connectivity and of the curing rates in the propagation dynamics.

\section*{Appendix}

\noindent
\textbf{Proof of Theorem \ref{th1}}.
Suppose that $\rho_i$ is any control. Suppose also that $w_i^*$ is the optimal disturbance obtained from the robust hamiltonian in (\ref{eq:H}), for given $\rho_i$. Thus
\begin{equation} \label{eq:proof1a}
\begin{array}{lll}
J^i_{x_i}(\rho_i, w_i, t) \\
\quad = \mathbb{E}_{i_t = i}^{\rho_i, w_i^*}\Bigg[\psi(i_T, x_{i_T}) + \int_t^T {i_{\tau}}(\tau)) - \frac{1}{2}\|w_{i_{\tau}}^*\|^2_{\Gamma_{i_{\tau}}}\Big] d\tau \Bigg] \\
\quad  = v_i(t) + \mathbb{E}_{i_t = i}^{\rho_i, w_i^*}\Bigg[\int_t^T \Big[\frac{dv_{i_{\tau}}}{dt}(\tau) + (A^{\rho}v)^{i_{\tau}}(\tau)  \\
\qquad  +g(i_{\tau}, x_{i_{\tau}}(\tau), \rho_{i_{\tau}}(\tau)) - \frac{1}{2}\|w_{i_{\tau}}^*\|^2_{\Gamma_{i_{\tau}}}\Big] d\tau \Bigg].
\end{array}
\end{equation}
The second equality above is obtained from the Dynkin formula
\begin{equation} \label{eq:Dynkin}
\begin{array}{lll}
\mathbb{E}_{i_t = i}^{\rho_i, w_i^*}\Big[v_{i_T}(T) - v_{i}(t)\Big] = \mathbb{E}_{i_t = i}^{\rho_i, w_i^*}\Bigg[\int_t^T \Big[\frac{dv_{i_{\tau}}}{dt}(\tau) \\
+ (A^{\beta}v)^{i_{\tau}}(\tau)d\tau\Bigg],
\end{array}
\end{equation}
where the terminal condition is $v_{i_T}(T) = \psi(i_T, x_{i_T})$ and $(A^{\rho}v)^{i_{\tau}}(\tau) = \sum_j \rho_{ij}(\tau)[v_j(\tau) - v_{i_{\tau}}(\tau)]$ is the infinitesimal generator of process $i_{\tau}$. Then we have that 
\begin{equation} \label{eq:proof1b}
\begin{array}{lll}
J^i_{x_i}(\rho_i, w_i, t) \ge v_i(t) + \mathbb{E}_{i_t = i}^{\rho_i, w_i^*}\Bigg[\int_t^T \Big[\frac{dv_{i_{\tau}}}{dt}(\tau) \\
\qquad + \min_{\mu(\cdot) \in (\mathbb{R}_0^+)^3} \sum_j \mu_j\Big[v_j(\tau) - v_{i_{\tau}}(\tau) \Big]  \\
\qquad  +g(i_{\tau}, x_{i_{\tau}}(\tau), \rho_{i_{\tau}}(\tau)) - \frac{1}{2}\|w_{i_{\tau}}^*\|^2_{\Gamma_{i_{\tau}}}\Big] d\tau \Bigg]. \\
\end{array}
\end{equation}
The above inequality follows from the fact that we minimise over any control $\mu(\cdot) \in (\mathbb{R}_0^+)^3$. From the definition of $w_i^*$, which is obtained by maximising over any possible disturbance,  we can rewrite the RHS in the above as follows 
\begin{equation} \label{eq:proof1c}
\begin{array}{lll}
J^i_{x_i}(\rho_i, w_i, t) \geq v_i(t) + \mathbb{E}_{i_t = i}^{\rho_i, w_i}\Bigg[\int_t^T \Big[\frac{dv_{i_{\tau}}}{dt}(\tau) \\
\qquad + \min_{\mu(\cdot) \in (\mathbb{R}_0^+)^3} \max_{w(\cdot) \in (\mathbb{R}_0^+)^3} \sum_j \mu_j\Big[v_j(\tau) - v_{i_{\tau}}(\tau) \Big] \\\qquad +g(i_{\tau}, x_{i_{\tau}}(\tau), \rho_{i_{\tau}}(\tau)) - \frac{1}{2}\|w_{i_{\tau}}\|^2_{\Gamma_{i_{\tau}}}\Big] d\tau \Bigg]. \\
\end{array}
\end{equation}
Replacing the minimax term by the robust Hamiltonian as in (\ref{eq:H}), we obtain:
\begin{equation} \label{eq:proof1d}
\begin{array}{lll}
J^i_{x_i}(\rho_i, w_i, t)  = v_i(t) + \mathbb{E}_{i_t = i}^{\rho_i, w_i^*}\Bigg[\int_t^T \Big[\frac{dv_{i_{\tau}}}{dt}(\tau) \\
\qquad + \mathcal{H}(x_{i_{\tau}}(\tau), \Delta_{i_{\tau}}v(\tau), i_{\tau}) \Big] d\tau \Bigg] \\
\quad  = v_i(t).
\end{array}
\end{equation}
To conclude the proof, the mean-field response as in (\ref{eq:HJ}) exists and it is the value function of the optimal control problem. By differentiating (\ref{eq:H}) with respect to $\rho_i$ and taking the gradient equal to zero, we have
\begin{equation} \label{eq:rhoL}
R_i\rho_i + \Delta_iv = 0,
\end{equation}
from which we obtain the optimal control in (\ref{eq:rho*}). Similarly,  by differentiating (\ref{eq:H}) with respect to $w_i$ and taking the gradient equal to zero, we have:
\begin{equation} \label{eq:wL}
-\Gamma_iw_i + \Delta_iv = 0,
\end{equation}
which yields the optimal adversarial disturbance as in (\ref{eq:w*}).\\

\noindent
\textbf{Proof of Theorem \ref{th2}}.
First we expand equation (\ref{eq:yi}) as 
\begin{equation} \label{eq:appeq}
\left\{\begin{array}{lll}
-\frac{1}{2}(\Gamma_{13}^{-1} + R_{31}^{-1})y_1^2 - \frac{1}{2}R_{32}^{-1}y_2^2+c_1 = \dot{y}_1,\\
-\frac{1}{2}R_{31}^{-1}y_1^2 - \frac{1}{2}(\Gamma_{23}^{-1} + R_{32}^{-1})y_2^2+c_2= \dot{y}_2,\\
\end{array}\right.
\end{equation}
where $c_1 = f_3(x_3) - f_1(x_1)$ and $c_2 = f_3(x_3) - f_2(x_2)$. From setting y dot 1 and y dot 2 equal we obtain:
\begin{equation} \label{eq:appeq2}
\begin{array}{lll}
-\frac{1}{2}(\Gamma_{13}^{-1} + R_{31}^{-1})y_1^2 - \frac{1}{2}R_{32}^{-1}y_2^2+c_1 = \\
\qquad \qquad = -\frac{1}{2}R_{31}^{-1}y_1^2 - \frac{1}{2}(\Gamma_{23}^{-1} + R_{32}^{-1})y_2^2+c_2,
\end{array}
\end{equation}
which can be simplified as
\begin{equation} \label{eq:appeq3}
-\frac{1}{2}\Gamma_{13}^{-1}y_1^2 +c_1 = -\frac{1}{2}\Gamma_{23}^{-1} y_2^2+c_2.
\end{equation}
From the last equation, we can now find the equilibrium points as in (\ref{eq:eqy}) when we use the values of $c_1,c_2$:
\begin{equation} \label{eq:appeq3}
y_1^2  = \Gamma_{23}^{-1} \Gamma_{13} y_2^2-2 \Gamma_{13} (f_2(x_2)-f_1(x_1)).
\end{equation}
We are interested in the negative values, which explains why we consider only the negative value of the square root. This concludes our proof.\\

\noindent
\textbf{Proof of Theorem \ref{th3}}.
Let us study the system around an equilibrium point for the distribution, which is given and constant over the horizon, i.e. $\mathbf{x}(t) = \mathbf{\hat{x}}: [0, T] \rightarrow \mathcal{S}^3$. We consider the following Jacobian for the system (\ref{eq:yi}): 
\begin{equation} \label{eq:jac2}
\begin{scriptsize}
J=\left[ \begin{array}{cc}
(\Gamma_{13}^{-1}+R_{31}^{-1})y_1  & R_{32}^{-1}y_2  \\
R_{31}^{-1}y_1 & (\Gamma_{23}^{-1}+R_{32}^{-1})y_2 \end{array} \right].
\end{scriptsize}
\end{equation}
Let us linearise the system and among the four equilibria represented by the 4 intersections between the two ellipses, the only asymptotically stable node is the one in the third quadrant. This derives from the fact that $\Delta > 0$ and $T^2 > 4\Delta$, where $\Delta$ is the determinant and $T$ is the trace of the Jacobian matrix.\\
The computation for the stability analysis include the calculation of the trace $T$ and determinant $\Delta$ of the Jacobian matrix in equation (\ref{eq:jac2}) at the equilibrium:
\begin{equation} \label{eq:appst}
\begin{array}{lll}
T = (\Gamma_{13}^{-1} + R_{31}^{-1})y_1 + (\Gamma_{23}^{-1} + R_{32}^{-1})y_2,\\
\Delta = (\Gamma_{13}^{-1} + R_{31}^{-1})(\Gamma_{23}^{-1} + R_{32}^{-1})y_1y_2 - R_{31}^{-1}R_{32}^{-1}y\sqrt{\xi}.\\
\end{array}
\end{equation}
We can easily expand $\Delta$, to see that it is always positive (which implies that there is no saddle point):
\begin{equation} \label{eq:appst2}
\Delta = (\Gamma_{13}^{-1}\Gamma_{23}^{-1} + \Gamma_{13}^{-1}R_{23}^{-1}R_{32}^{-1} + R_{31}^{-1}\Gamma_{23}^{-1})y_1y_2>0.\\
\end{equation}
Analogously, we compute the square of the trace and compare it to the determinant to see if we have an asymptotically stable or unstable node:
\begin{equation} \label{eq:appst3}
\begin{array}{lll}
T^2 =(\Gamma_{13}^{-1} + R_{31}^{-1})^2y_1^2 + (\Gamma_{23}^{-1} + R_{32}^{-1})^2y_2^2 +\\
\qquad \qquad+ 2(\Gamma_{13}^{-1} + R_{31}^{-1})(\Gamma_{23}^{-1} + R_{32}^{-1})^2y_1y_2.
\end{array}
\end{equation}
From the above equation, it is easy to see that $T^2 > 4\Delta$ for the third quadrant, then we have an asymptotically stable node.
This concludes our proof.\\

\noindent
\textbf{Proof of Theorem \ref{th4}}.
The proof for each statement follows:
\begin{enumerate}
\item The fact that, if $s(0), r(0) \in [0,1]_n$, then $s(t), r(t) \in [0,1]_n$ for all $t>0$, means that $[0,1]_n$ is an invariant set for the differential equation (\ref{eq:CDMvf1}). This can be seen by specializing the right-hand side of both equations in (\ref{eq:CDMvf1}) in $0$ and $1$ and check whether the derivatives are positive and negative definite, respectively. To see this, note that, in $0$, we have $\dot{s}(t) = -\beta_{23}' diag(\bar 0_n)Ar(t) + \beta_{32}' diag(z(t))A\bar 0_n -\beta_{23}'' \bar 0_n + \beta_{32}'' z(t) = \beta_{32}'' z(t)$, and $\dot{r}(t) = -\beta_{13}' diag(\bar 0_n)As(t) + \beta_{31}' diag(z(t))A\bar 0_n -\beta_{13}'' \bar 0_n + \beta_{31}'' z(t) = \beta_{31}'' z(t)$, which are positive definite. Analogously, in $1$, we have $\dot{s}(t) = -\beta_{23}' diag(\bar 1_n)A\bar 0_n + \beta_{32}' diag(\bar 0_n)A\bar 1_n -\beta_{23}'' \bar 1_n + \beta_{32}'' \bar 0_n = -\beta_{23}''$, and $\dot{r}(t) = -\beta_{13}' diag(\bar 1_n)A\bar 0_n + \beta_{31}' diag(\bar 0_n)A\bar 1_n -\beta_{13}'' \bar 1_n + \beta_{31}'' \bar 0_n = -\beta_{13}''$, which are negative definite. Then, for the property of continuity of the equations, we can conclude that $[0,1]_n$ is an invariant set for the set of differential equations (\ref{eq:CDMvf1}).\\

\item Let us rewrite system (\ref{eq:CDMvf1}) by removing the explicit dependance on time as:
\begin{equation} \label{eq:CDMvf2}
\left\{\begin{array}{lll}
\dot{s} = -\beta_{23}' diag(s)Ar + \beta_{32}' diag(\bar 1_n - s - r)As\\
\qquad \quad -\beta_{23}'' s + \beta_{32}'' (\bar 1_n - s - r), \\
\dot{r} = -\beta_{13}' diag(r)As + \beta_{31}' diag(\bar 1_n - s - r)Ar\\
\qquad \quad -\beta_{13}'' r + \beta_{31}'' (\bar 1_n - s - r), \\
\end{array}\right.
\end{equation}
where we have $z = \bar 1_n - s - r$, by taking into account the conservation of mass.
By inspection, it is straightforward to prove that the pairs $(\bar 1_n, \bar 0_n)$ and $(\bar 0_n, \bar 1_n)$ are equilibrium points of the above system.\\
To prove stability, consider system (\ref{eq:CDMvf2}). The corresponding Jacobian matrix linearized about $(\bar 1_n, \bar 0_n)$ is
\begin{equation} \label{eq:jacCDM10}
\begin{scriptsize}
J=\left[ \begin{array}{cc}
-\beta_{23}'' \bar 1_n - \beta_{32}'' \bar 1_n & -\beta_{23}'A \bar 1_n -\beta_{32}'A \bar 1_n - \beta_{32}'' \bar 1_n \\
-\beta_{31}'' \bar 1_n & -\beta_{13}'A \bar 1_n - \beta_{13}'' \bar 1_n - \beta_{31}'' \bar 1_n \end{array} \right],
\end{scriptsize}
\end{equation}
and the Jacobian matrix linearized about $(\bar 0_n, \bar 1_n)$ is
\begin{equation} \label{eq:jacCDM01}
\begin{scriptsize}
J=\left[ \begin{array}{cc}
-\beta_{23}'A \bar 1_n - \beta_{23}'' \bar 1_n - \beta_{32}'' \bar 1_n & - \beta_{32}'' \bar 1_n \\
-\beta_{13}'A \bar 1_n -\beta_{31}'A \bar 1_n \beta_{31}'' \bar 1_n & - \beta_{13}'' \bar 1_n - \beta_{31}'' \bar 1_n \end{array} \right].
\end{scriptsize}
\end{equation}
For both matrices, the trace is negative. We can ensure asymptotic stability when the coefficients of the main diagonal of the Jacobian matrix are, in absolute value, greater than those in the off-diagonal. Otherwise, we have saddle points.\\

\item We now prove that the pair $((1/(2+k))_n, (1/(2+k))_n)$ is a set of equilibrium points. We seek equilibrium points that are symmetric, i.e. $s=r$. We can rewrite the first equation of (\ref{eq:CDMvf2}) as
\begin{equation} \label{eq:CDMeq}
\begin{array}{lll}
\dot{s} = -(\beta_{23}'+2\beta_{32}') diag(s)As + \beta_{32}'As \\
\qquad \quad -\beta_{23}''s + \beta_{32}'' (\bar 1_n - 2s).
\end{array}
\end{equation}
By linearizing about $((1/(2+k))_n, (1/(2+k))_n)$, by inspection, we have
\begin{equation} \label{eq:CDMeq2}
\begin{array}{lll}
\dot{s} = -(k\beta_{32}'+2\beta_{32}') diag((1/(2+k))_n)A(1/(2+k))_n \\
+ \beta_{32}'A(1/(2+k))_n +\beta_{32}''(\bar 1_n - (2 + k) (1/(2+k))_n),
\end{array}
\end{equation}
where the first two terms and the last one cancel out. Thus, $((1/(2+k))_n, (1/(2+k))_n)$ is an equilibrium point of the system. For system (\ref{eq:CDMeq}), $((1/(2+k))_n, \bar 1_n - (2/(2+k))_n, (1/(2+k))_n)$, is the corresponding set of equilibrium points.\\
To prove stability, consider system (\ref{eq:CDMvf2}). The corresponding Jacobian matrix linearized about $((1/(2+k))_n, (1/(2+k))_n)$ is
\begin{equation} \label{eq:jacCDM2}
\begin{scriptsize}
J=\left[ \begin{array}{cc}
-\beta_{32}'A \bar 1_n - \beta_{32}'' \bar 1_n & -\beta_{32}'A \bar 1_n - \beta_{32}'' \bar 1_n \\
-\beta_{31}'A \bar 1_n - \beta_{31}'' \bar 1_n & -\beta_{31}'A \bar 1_n - \beta_{31}'' \bar 1_n \end{array} \right].
\end{scriptsize}
\end{equation}
The determinant of (\ref{eq:jacCDM2}) is $\Delta(J) = 0$, while the trace is $Tr(J) = -\beta_{32}'A \bar 1_n - \beta_{32}'' \bar 1_n -\beta_{31}'A \bar 1_n - \beta_{31}'' \bar 1_n$. So, if $Tr(J)<0$ the equilibrium point for the system, i.e. $(s^*, \bar 0_n, \bar 1_n - s^*)$, is asymptotically stable. This is true since all the coefficients of the Jacobian matrix are negative.\\
\end{enumerate}

\noindent
\textbf{Proof of Theorem \ref{th5}}.
Each statement is proved in the following:
\begin{enumerate}
\item If $s(0), r(0) \in [0,1]_n$, then $s(t), r(t) \in [0,1]_n$ for all $t>0$, it means that $[0,1]_n$ is an invariant set for the differential equation (\ref{eq:VPNvf1}). By setting the equations in $0$ and $1$, it must be checked whether the derivatives are positive and negative definite, respectively. So, in $0$, we have $\dot{s}(t) = -\beta_{23}diag(\bar 0_n)Az(t) + \beta_{32} z(t) = \beta_{32} z(t)$, and $\dot{r}(t) = -\beta_{13}diag(\bar 0_n)Az(t) + \beta_{31} z(t) = \beta_{31} z(t)$, which are positive definite. Similarly, in $1$, we have $\dot{s}(t) = -\beta_{23}diag(\bar 1_n)A\bar 0_n + \beta_{32} \bar 0_n = 0$, and $\dot{r}(t) = -\beta_{13}diag(\bar 1_n)A\bar 0_n + \beta_{31} \bar 0_n = 0$, which are nonpositive definite. Since the equations are continuous, then we can conclude that $[0,1]_n$ is an invariant set for the set of differential equations (\ref{eq:VPNvf1}).\\
\item Let us rewrite system (\ref{eq:VPNvf2}) by removing the explicit dependance on time and by taking into account the conservation of mass as:
\begin{equation} \label{eq:VPNvf3}
\left\{\begin{array}{lll}
\dot{s} = -\beta_{23}diag(s^*)A(\bar 1_n - s - r) + \beta_{32} (\bar 1_n - s - r), \\
\dot{r} = -\beta_{13}diag(\bar 1_n - s^*)A(\bar 1_n - s - r) + \beta_{31} (\bar 1_n - s - r). \\
\end{array}\right.
\end{equation}
The corresponding Jacobian matrix is therefore:
\begin{equation} \label{eq:jacVPN}
\begin{scriptsize}
J=\left[ \begin{array}{cc}
\zeta & \zeta \\
\eta & \eta \end{array} \right],
\end{scriptsize}
\end{equation}
where $\zeta = \beta_{23}diag(s^*)A \bar 1_n - \beta_{32} \bar 1_n$ and $\eta = \beta_{13}diag(\bar 1_n - s^*)A \bar 1_n - \beta_{31} \bar 1_n$.
The determinant of (\ref{eq:jacVPN}) is $\Delta(J) = 0$, while the trace is $Tr(J) = \beta_{23}diag(s^*)A \bar 1_n - \beta_{32} \bar 1_n + \beta_{13}diag(\bar 1_n - s^*)A \bar 1_n - \beta_{31} \bar 1_n$. So, if $Tr(J)<0$ the equilibrium point for the system, i.e. $(s^*, \bar 0_n, \bar 1_n - s^*)$, is asymptotically stable. This holds true when the condition (\ref{eq:VPNc}) holds.

\item When $\lim_{\beta_{32},\beta_{13} \rightarrow 0}$, system (\ref{eq:VPN1}) can be rewritten in vector form as
\begin{equation} \label{eq:2SIR}
\left\{\begin{array}{lll}
\dot{s}(t) = -\beta_{23}diag(s(t))Az(t), \\
\dot{z}(t) = \beta_{23}diag(s(t))Az(t) - \beta_{31} z(t), \\
\dot{r}(t) = \beta_{31} z(t), \\
\end{array}\right.
\end{equation}
which is the equivalent of the standard $SIR$ network model.
\end{enumerate}

\begin{IEEEbiography}
[{\includegraphics[width=1in,height=1.25in,clip,keepaspectratio]{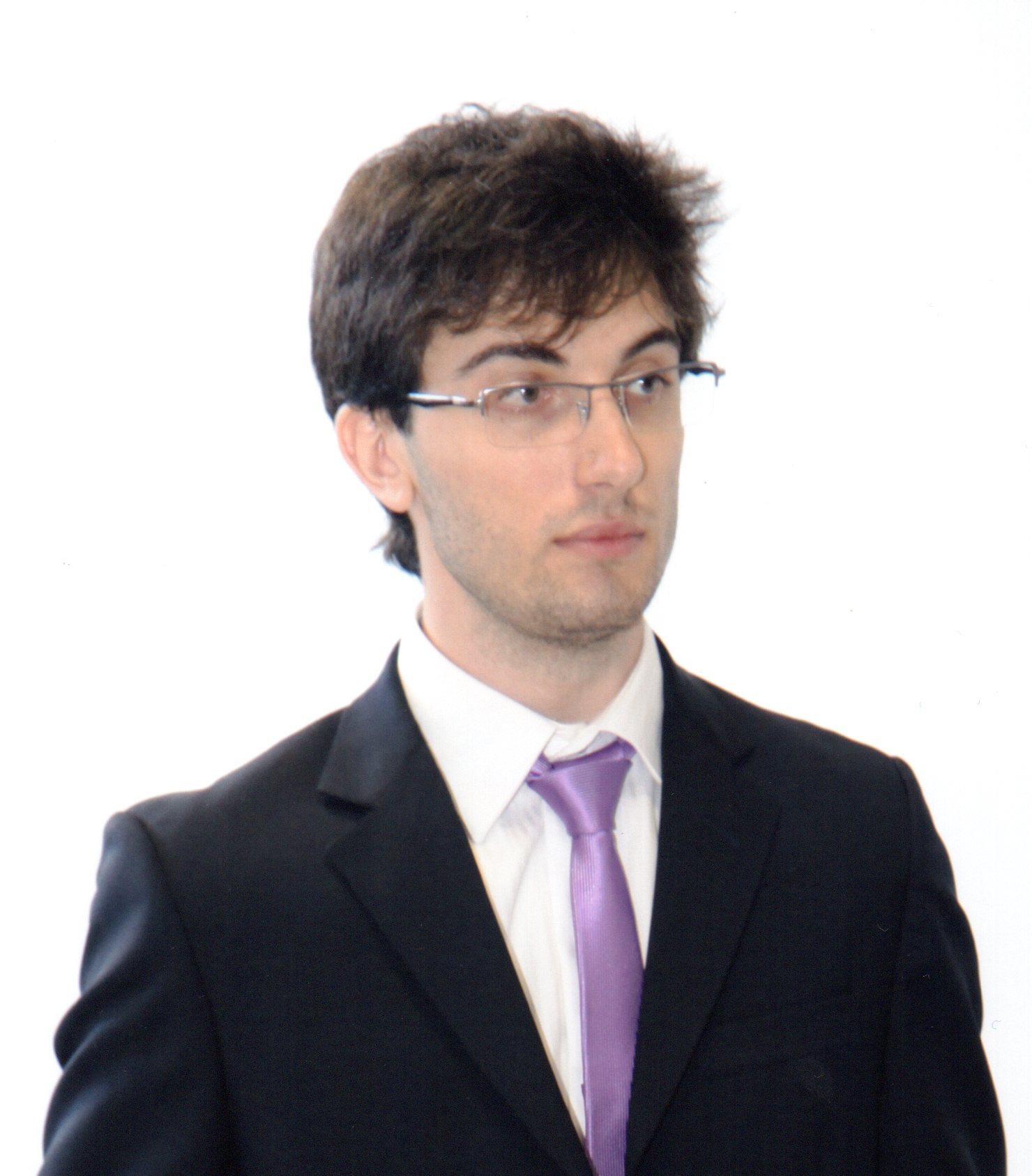}}]
{Leonardo Stella} graduated from the University of Palermo, Italy, in Computer Engineering with his final year dissertation on the subject of ``Opinion dynamics and stubbornness through mean-field games'' in 2013, after spending one year Bachelor at USI, Universit\`a della Svizzera Italiana, Switzerland, in Computer Science. He received his master degree in 2016 from University Sapienza, Italy, in Artificial Intelligence and Robotics with his final year dissertation working jointly with the team of University Tor Vergata for the EU project ``Superfluidity''. His main interests include Artificial Intelligence, Complexity, Game Theory, Big Data, Multi-Agent Systems, Natural Language Processing and their applications. 

He is conducting PhD research into 'Optimal Control of Intelligent Agents for Collective Decision Making in
Biologically Inspired and AI Models' under the supervision of Dr Dario Bauso and Dr Roderich Gross in the Department of Automatic Control and Systems Engineering, University of Sheffield.
\end{IEEEbiography}

\begin{IEEEbiography}[{\includegraphics[width=1in,height=1.25in,clip,keepaspectratio]{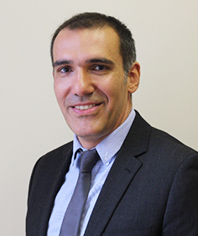}}]
{Dario Bauso} received the Laurea degree in Aeronautical Engineering in 2000 and the Ph.D. degree in Automatic Control and System Theory in 2004 from the University of Palermo, Italy. Since 2015 he has been with the Department of Automatic Control and Systems Engineering, The University of Sheffield (UK), where he is currently Reader in Control and Systems Engineering. Since 2005 he has also been with the Dipartimento di Ingegneria Chimica, Gestionale, Informatica, Meccanica, University of Palermo (Italy), where he is currently Associate Professor of Operations Research.  From 2012 to 2014 he was also Research Fellow at the Department of Mathematics, University of Trento (Italy).

He has been academic visitor in several universities. From October 2001 to June 2002, he was a Visiting Scholar at the Mechanical and Aerospace Engineering Department, University of California, Los Angeles (USA). In 2010 he was short-term visiting scholar at the Department of Automatic Control of Lund University (Sweden) and at the Laboratory of Information and Decision Systems of the Massachusetts Institute of Technology (USA). In 2013 he was visiting lecturer at the Department of Engineering Science, University of Oxford (UK) and at the Department of Electrical and Electronic Engineering of Imperial College London (UK).

His research interests are in the field of Optimization, Optimal and Distributed Control, and Game Theory. Since 2010 he is member of the Conference Editorial Board of the IEEE Control Systems Society. He is Associate Editor of Dynamic Games and Applications since 2011, and of IEEE Transactions on Automatic Control since 2013. He has also been general chair of the 6th Spain, Italy, and Netherlands Meeting on Game Theory (SING~6).
\end{IEEEbiography}

\end{document}